\documentclass[aps,prd,twocolumn,groupedaddress, showkeys, showpacs]{revtex4}
\usepackage{graphicx,subfigure,enumerate,pstricks,latexsym,longtable,amsmath,amssymb}
\usepackage[english]{babel}

\newcommand{\beq}{\begin{equation}} \newcommand{\eeq}{\end{equation}}
\newcommand{\bea}{\begin{eqnarray}} \newcommand{\eea}{\end{eqnarray}}
\def\nn{\nonumber}
\providecommand{\dif}{\mathrm{d}} \def\d{\dif}
\newcommand{\Schw}{Schwarzschild}

\newcommand{\ce}{{\cal{E}}} \newcommand{\cl}{{\cal{L}}}  \newcommand{\cb}{{\cal{B}}}  

\def\p{p} 
\def\cp{\pi} 

\def\af{\zeta} 

\def\HP{\widetilde{H}} 

\def\RS{R} \def\DD{\Delta} \def\QS{Q} \def\SS{S} 

\def\mJ{{\rm I}} 
\def\mD{{\rm II}} 

\graphicspath{{pic/}}

\begin{document}

\title{Acceleration of charged particles due to chaotic scattering in the combined black hole gravitational field and asymptotically uniform magnetic field}

\author{Zden\v{e}k Stuchl{\'i}k}
\email{zdenek.stuchlik@fpf.slu.cz}
\author{Martin Kolo\v{s}}
\email{martin.kolos@fpf.slu.cz}
\affiliation{Institute of Physics and Research Centre of Theoretical Physics and Astrophysics, Faculty of Philosophy and Science, Silesian University in Opava, \\
Bezru{\v c}ovo n{\'a}m.13, CZ-74601 Opava, Czech Republic}

\begin{abstract}
To test the role of large-scale magnetic fields in accretion processes, we study dynamics of charged test particles in vicinity of a black hole immersed into an asymptotically uniform magnetic field. 
Using the Hamiltonian formalism of charged particle dynamics, we examine chaotic scattering in the effective potential related to the black hole gravitational field combined with the uniform magnetic field. Energy interchange between the translational and oscillatory modes od the charged particle dynamics provides mechanism for charged particle acceleration along the magnetic field lines. This energy transmutation is an attribute of the chaotic charged particle dynamics in the combined gravitational and magnetic fields only, the black hole rotation is not necessary for such charged particle acceleration. The chaotic scatter can cause transition to the motion along the magnetic field lines with small radius of the Larmor motion or vanishing Larmor radius, when the speed of the particle translational motion is largest and can be ultra-relativistic. 
We discuss consequences of the model of ionization of test particles forming a neutral accretion disc, or heavy ions following off-equatorial circular orbits, and we explore the fate of heavy charged test particles after ionization where no kick of heavy ions is assumed and only switch-on effect of the magnetic field is relevant. We demonstrate that acceleration and escape of the ionized particles can be efficient along the Kerr black hole symmetry axis parallel to the magnetic field lines. We show that strong acceleration of ionized particles to ultra-relativistic velocities is preferred in the direction close to the magnetic field lines. Therefore, the process of ionization of Keplerian discs around Kerr black holes can serve as a model of relativistic jets.
\end{abstract}

\pacs{04.70.Bw, 04.25.-g, 04.70.-s, 97.60.Lf \hfill
}

\maketitle

\section{Introduction}

In the processes occurring around black holes the magnetic fields can be relevant due to several reasons. The local magnetic fields in the Keplerian accretion discs are assumed to be the source of the basic viscosity mechanism of accretion due to the magneto-rotational instability \cite{Bal-Haw:1991:APJ:}. The kinetic dynamo effect in collisionless plasma in accretion discs can create global toroidal magnetic fields \cite{Cre-Stu:2013:PHYSRE:}. Many of the observed black hole candidates are assumed to have an accretion disc constituted from conducting plasma which dynamics can generate a regular magnetic field. The kinetic effects of collisionless plasmas could generate equilibrium configurations of plasmas in various conditions under combined gravitational and magnetic fields \cite{Cre-Tes-Mil:2012:PHYSRL:,Cre-Stu-Tes:2013:PhysPlasm:,Cre-Tes-Stu:2014:PhysPlasm:} or could govern transitions from neutral to ionized equilibria of accretion discs \cite{Cre-Stu:2014:PhysPlasm:}

\begin{figure*}
\includegraphics[width=\hsize]{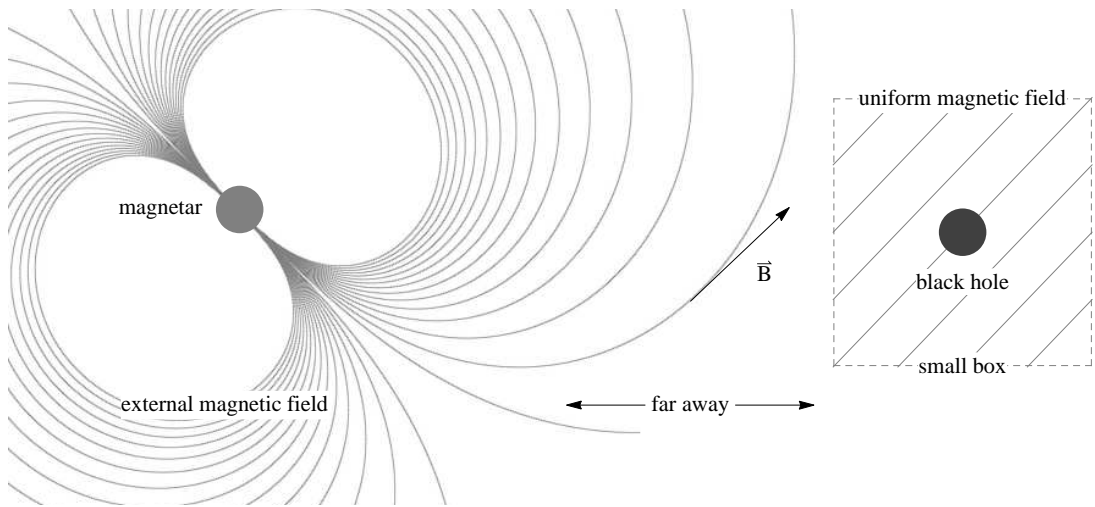}
\caption{ \label{motivace}
Black hole immersed in an external electromagnetic field. Large scale electromagnetic field can have dipole character for a magnetar, but at large distance from the source its character can be simplified to almost uniform magnetic field in finite element of space as shown in \cite{Kov-etal:2014:PHYSR4:}.
}
\end{figure*}

If a rotating black hole carries by itself an electric charge, being described by the Kerr-Newman background, it has its intrinsic electromagnetic field that could influence dynamics of charged particles in accreting matter \cite{Mis-Tho-Whe:1973:Gra:,Ruf:1973:BlaHol:}. The motion equations of the charged test particles are then separable and integrable and the motion has regular character \cite{Ali-Gal:1981:GRG:,Bic-Stu-Bal:1989:BAC:,Bal-Bic-Stu:1989:BAC:,Stu-Bic-Bal:1999:GRG:,Pug-Que-Ruf:2011:PHYSR4:,Pug-Que-Ruf:2013:PHYSR4:}. 

The physical processes in the surrounding of black holes could be influenced also by large scale magnetic fields not related directly to the black hole. Such magnetic fields could be of cosmological origin \cite{Gra-Rub:2001:PHYSREP:,Dur-Ner:2013:AAPR:,Kro:1994:RPP:,Ner-Vov:2010:SCI:,Rat:1992:APJL:,Jai-Slo:2012:PRD:} or they could be related to some source that can demonstrate a complex structure in vicinity of field source, but at large distances, their character can be simple and close to a homogeneous magnetic field (see Fig. \ref{motivace}.) - for simplicity, such magnetic fields are considered to be asymptotically uniform as discussed in \cite{Wald:1974:PHYSR4:}. The motion in the gravitational field of a black hole combined with an external electromagnetic field is not separable and has in general chaotic character. There is a large variety of studies of the charged test particle motion in such combined fields \cite{Prasanna:1980:RDNC:,Ali-Gal:1981:GRG:,Fro-Sho:2010:PHYSR4:,Pre:2004:PHYSR4:,Abd-etal:2013:PHYSR4:,Bak-etal:2010:CLAQG:,Kop-etal:2010:APJ:,Kol-Stu-Tur:2015:CLAQG:}. The energy from collision of charged particles in the vicinity of the horizon of a black hole or a naked singularity immersed into an external magnetic field can cause particle acceleration \cite{Fro:2012:PHYSR4:,Zah-etal:2013:PHYSR4:,Stu-Sche-Abd:2014:PHYSR4:,Shi-Kim-Chi:2014:PHYSR4:}.  Of special interest is existence of off-equatorial circular orbits of charged particles \cite{Kov-Stu-Kar:2008:CLAQG:,Kov-etal:2010:CLAQG:} and related existence of toroidal charged-fluid configurations levitating off the equatorial plane \cite{Sla-etal:2013:ApJS:,Cre-etal:2013:ApJS:,Kov-etal:2014:PHYSR4:}. 

Recently, it has been found that the center of the Galaxy has a strong magnetic field around a supermassive black hole that is not related to an accretion disc \cite{Eat-etal:2013:NATUR:}. Therefore, the possibility that black holes can be immersed in an external, large scale electromagnetic field has to be taken quite seriously. Moreover, it has been demonstrated that a black hole located near the equatorial plane of a magnetar will be immersed in a nearly homogeneous magnetic field if distance to the magnetar is large enough \cite{Kov-etal:2014:PHYSR4:}. Hereafter in this paper we will concentrate our attention on the particular and simplified case of a black hole immersed in an asymptotically uniform magnetic field known as Wald solution for a magnetized black hole \cite{Wald:1974:PHYSR4:}. We shall study dynamics of charged test particles in the combined gravitational and electromagnetic fields of such configurations assuming for simplicity that the symmetry of the black hole spacetime is in accord with the symmetry of the asymptotically uniform magnetic field, i.e., the symmetry axis of the Kerr spacetime is aligned with the field lines of the uniform magnetic field. Then the motion equations of charged test particles allow for existence of motion constants (energy and axial angular momentum) simplifying thus significantly treatment of the motion at one side, and keeping the relevant signatures of the interplay of the gravitational and magnetic fields on the other side.

The motion of neutral test particles is not influenced by magnetic fields satisfying the condition of the test field approximation, $B<<10^{19}{M}_{\odot}/{M}$~Gauss, that is satisfied even in close vicinity of magnetars. However, the motion of charged test particles in close vicinity of a black hole horizon could be strongly influenced even by relatively weak test magnetic fields \cite{Kol-Stu-Tur:2015:CLAQG:}. For a charged test particle with charge $q$ and mass $m$ moving in vicinity of a black hole with mass $M$ surrounded by an uniform magnetic field of the strength $B$, one can introduce a dimensionless quantity $b$ that can be identified as relative Lorenz force \cite{Fro-Sho:2010:PHYSR4:} $b=|q| B G M / m c^4 $. This quantity can be really quite large even for weak magnetic fields due to the large value of the specific charge $q/m$. 

Recently a variety of phenomena related to the combined gravo-magnetic effect on charged test particle motion has been studied. The study of charged particles 'kicked' from the innermost stable circular orbit (ISCO) in equatorial plane and hence escaping to infinity along the axis of symmetry has been treated in \cite{Shi-Kim-Chi:2014:PHYSR4:} -- because the charged particle motion in the vicinity of a black hole immersed into magnetic field is chaotic, the resulting final ejection velocity does not depend continuously on the initial conditions \cite{Kop-etal:2010:APJ:}. It has been demonstrated that collisions of particles in vicinity of black holes could be enhanced by the external magnetic fields \cite{Iga-Har-Kim:2012:PHYSR4:,Abd-etal:2013:AstrSpaSci:,Zas:2014:MPLA:,Tos-Ahm-Abd-Stu:2014:PHYSR4:,Zas:2015:EPJC:}, but these processes can be observationally efficient especially in the field of Kerr naked singularities \cite{Stu-Sche:2013:CLAQG:}. Relation of the quasi-circular equatorial motion of charged test particles around black holes immersed in the magnetic field to the high-frequency quasi-period oscillations observed in some microquasars has been demonstrated in \cite{Kol-Stu-Tur:2015:CLAQG:}. Synchrotron radiation of the charged particles following quasi-circular orbits in the combined gravitational and electromagnetic fields has been studied in \cite{Ali-Gal:1981:GRG:,Sho:2015:ArXiv:}. 

The energy of a rotating black hole immersed into a magnetic field can be extracted due to the Blandford-Znajek process \cite{Bla-Zna:1977:MNRAS:} and demonstrated by relativistic jets, i.e., collimated streams of particles escaping the central object along the axis of rotation with relativistic velocities.
In the present paper we explore the mechanism hidden behind the charged particle ejection using the theory of chaotic scattering in the combined effective potential of the black hole and the asymptotically uniform magnetic field. Energy of the charged particle in such combined fields can be separated into two modes, one related to the direction along the magnetic field lines, the other to the perpendicular directions. We are able to demonstrate that an energy transmission mechanism, i.e., interchange between the two energy modes od the charged particle dynamics due to chaotic scattering in the deep gravitational field near the black hole horizon, can provide sufficient energy for ultra-relativistic motion of charged particle along the magnetic field lines.
As a source of charged particles, we explore a model of ionization of test particles forming a neutral accretion disk, where no 'kick' is needed for a charged particle (e.g. a heavy ion or proton) to leave the circular or quasi-circular orbit. The ionization can be realized by an irradiation of a part of the originally neutral disc when an electron carries the "kick" of the irradiation, while the heavy ion only feels switched-on Lorentz force. Moreover, heavy ions following off-equatorial orbits in the combined gravo-magnetic fields \cite{Kov-Stu-Kar:2008:CLAQG:,Kov-etal:2010:CLAQG:} could be also accelerated in the fields as an irradiation of such a heavy ion can increase its specific charge due to increasing degree of ionization. We expect that our introductory study can be of relevance for understanding of various astrophysical phenomena observed in the systems where compact objects (black holes, naked singularities, neutron stars, quark stars) with strong gravity are assumed. 

Throughout the paper, we use the spacetime signature $(-,+,+,+)$, and the system of geometric units in which $G = 1 = c$. However, for expressions having an astrophysical relevance we use the speed of light explicitly. Greek indices are taken to run from 0 to 3.

\section{Charged test particle dynamics}

We use the Hamiltonian formulation for dynamics of charged test particles with specific charge $q/m$ in the vicinity of the axially symmetric black hole immersed in an external asymptotically uniform magnetic field. The dynamics of the neutral test particle motion governed by the geodesic structure of the black hole geometry can be obtained by puting $q=0$ in the Hamiltonian formalism.

Kerr black holes are described by the Kerr geometry that is given in the standard Boyer-Lindquist coordinates and the geometric units in the form
\bea
 \d s^2 &=& - \left( 1- \frac{2Mr}{\RS^2} \right) \d t^2 - \frac{4Mra \sin^2\theta}{\RS^2} \, \d t \d \phi \nonumber\\
 && + \left( r^2 +a^2 + \frac{2Mra^2}{\RS^2} \sin^2\theta \right) \sin^2\theta \, \d \phi^2 \nonumber \\
 && + \frac{\RS^2}{\DD} \, \d r^2 + \RS^2\, \d\theta^2, 
 \label{KerrMetric} 
\eea
where
\beq
\RS^2 = r^2 + a^2 \cos^2\theta, \quad \DD = r^2 - 2Mr + a^2, \label{RSaDD}
\eeq
$a$ denotes spin and $M$ gravitational mass of the spacetimes that fulfil condition $a \leq M$ for black holes, and $a>M$ in the naked singularity case. The physical singularity is located at the ring $r=0, \theta = \pi/2$ that can be well characterized in the so called Kerr-Schild "Cartesian" coordinates that are related to the Boyer-Lindquist coordinates by the relations \cite{Car:1973:BlaHol:}
\bea
x &=& (r^2+a^2)^{1/2}\sin \theta\cos\left[\phi-\tan^{-1}\left(\frac{a}{r}\right)\right], \label{cordKS1}\\
y &=& (r^2+a^2)^{1/2}\sin \theta\sin\left[\phi-\tan^{-1}\left(\frac{a}{r}\right)\right], \label{cordKS2}\\
z &=& r\cos\theta. \label{cordKS3}
\eea

Because of the axial symmetry we are interested only in constant $\phi$ sections of the whole spacetime; we are free to choose 
\beq
\phi = \tan^{-1}\left(a/r\right)
\eeq
obtaining coordinate transformation in $r,\theta$ ($x,z$) plane
\beq
 x = \sqrt{r^2 + a^2} \sin \theta ,\quad z = r \cos \theta. \label{xycord}
\eeq 
At the $x$--$y$ plane, the physical singularity is located at $x=\pm a$ and $z=0$.

The Kerr metric (\ref{KerrMetric}) is asymptotically flat, i.e. far away from the black hole ($r \to \infty$), the Kerr metric becomes to be Minkowski flat metric. The Kerr asymptotic limit can be obtained by taking $M=0$ in (\ref{KerrMetric}) and rewriting the metric using the Kerr-Schild coordinates (\ref{cordKS1}-\ref{cordKS3}) into manifestly flat "Cartesian" $(t,x,y,z)$ form 
\beq
 \d s^2 = -\d t^2 + \d x^2 + \d y^2 + \d z^2. \label{KerrFlat} 
\eeq
We can also use the cylindrical coordinates $(t,\rho,\phi,z)$ and rewrite (\ref{KerrFlat}) in the form 
\beq
 \d s^2 = -\d t^2 + \d \rho^2 + \rho^2 \d \phi^2 + \d z^2, \label{KerrCyli} 
\eeq
where new coordinates $\rho, \phi$ are given by $\rho^2=x^2+y^2$ and $\phi=\arctan(y/x)$. 

In the following, we put $M=1$, i.e., we use dimensionless radial coordinate $r$ and dimensionless spin $a$, or dimensionless time coordinate $t$. There is no event horizon in the naked singularity spacetimes, in contrast to the Kerr black hole spacetimes (with $a<1$) when two event horizons exist. In the present paper we restrict our attention to the black hole spacetime regions located above the outer event horizon at $r_{+}=1+(1-a^2)^{1/2}$. 

\begin{figure*}
\subfigure[~ $Q=0$]{\label{figVeff1} \includegraphics[width=0.47\hsize]{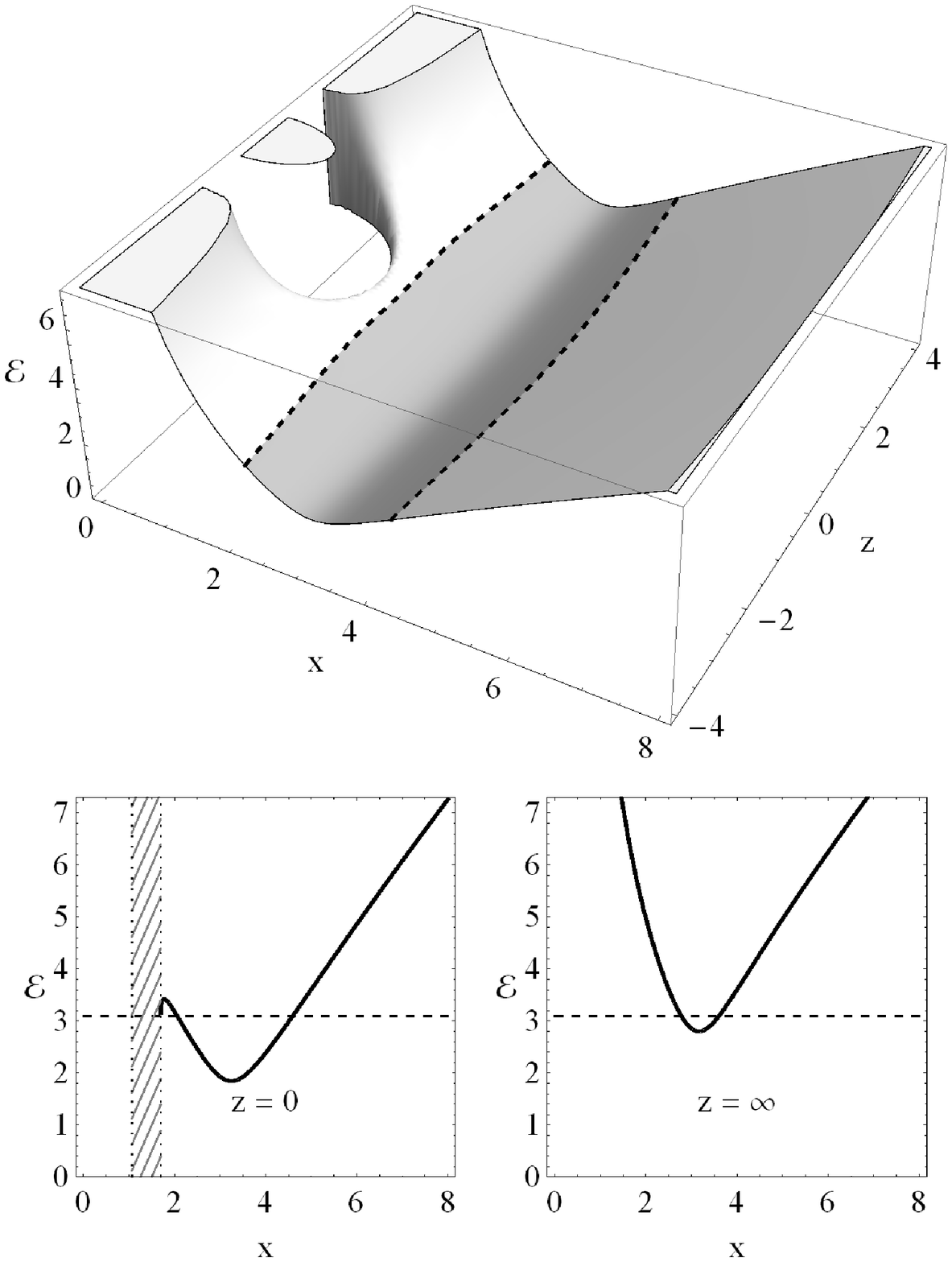}}
\subfigure[~ $Q=Q_{\rm W}$]{\label{figVeff2} \includegraphics[width=0.47\hsize]{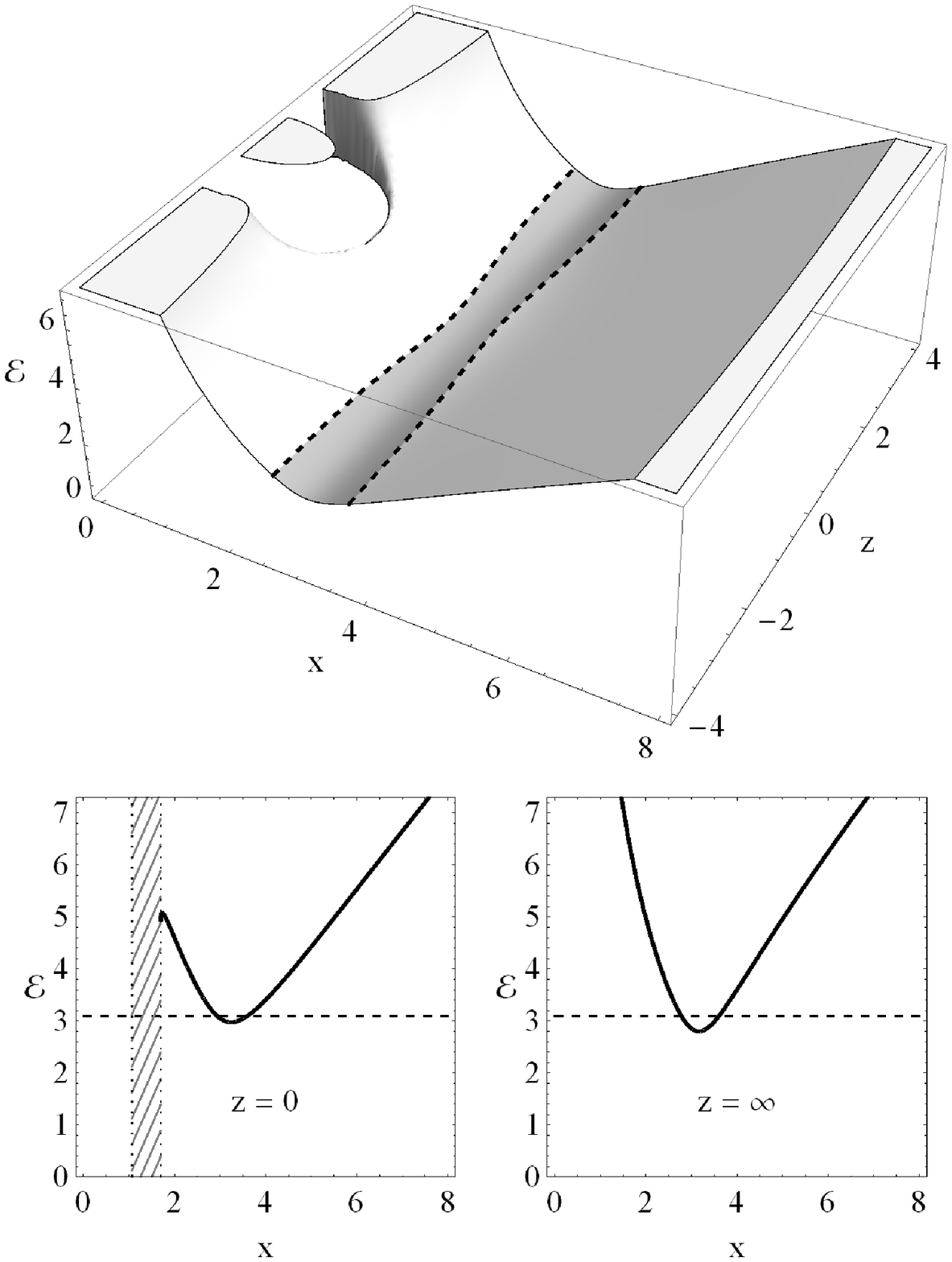}}
\caption{
\label{figVeff}
The effective potential $V_{\rm eff}(x,z;a,\cb,\cl)$ (\ref{VeffCharged}) for motion of charged particles in the combined gravitational field of Kerr black hole with $a=0.9$ and the uniform magnetic field with $\cb=1$. We give typical examples of the effective potential $V_{\rm eff}(x,y)$ for angular momentum $\cl=10$; we also present sections of $V_{\rm eff}$ in the equatorial plane, $z=0$, and at infinity, $z=\infty$. For the charged particle with energy $\ce=3.1$, the energy boundary for the particle motion (dashed curves given by the condition $\ce=V_{\rm eff}$) is open to infinity, allowing the charged particle to escape to infinity along the $z$-axis.
The effective potential $V_{\rm eff}$ for both cases of neutral black hole, $Q=0$, and the black hole with the Wald charge $Q=Q_{\rm W}$ are presented. For both cases the effective potential $V_{\rm eff}$ coincide at $z=\infty$, the main difference between them occurs in the equatorial plane, where the energetic boundary is widest for the $Q=0$ case, while it is narrowest for the $Q=Q_{\rm W}$ case.
The influence of the black hole  gravitational field is important close to the black hole horizon $x^2+z^2\rightarrow~r_{+}^2$ where the $V_{\rm eff}$ is diverging - $V_{\rm eff}(x,y)$ is not defined between inner and outer horizons. The influence of the external magnetic field becomes important at large values of coordinates $x$ and $z$.
}
\end{figure*}

For the external weak asymptotically uniform magnetic field, with the magnetic field vector $\vec{B}$ parallel to the spacetime axis of symmetry at $z$, the electromagnetic four-vector potential $A^{\alpha}$ takes the form \cite{Wald:1974:PHYSR4:}
\bea
A_{t} &=& \frac{B}{2} (g_{t\phi} + 2 a g_{tt}) - \frac{Q}{2} g_{tt} - \frac{Q}{2}, \\ \label{emAt}
A_{\phi} &=& \frac{B}{2} (g_{\phi\phi} + 2 a g_{t\phi}) - \frac{Q}{2} g_{t\phi}, \label{emAp}
\eea
where $B$ is magnitude of the asymptotically homogeneous magnetic field and $Q$ is the charge of the black hole related to the charging of the black hole due to the magnetic field influence \cite{Wald:1974:PHYSR4:}. The electromagnetic field is weak in the sense it is not contributing to the geometry of spacetime, but it can still have strong influence on the charged particle dynamics.
As show by Wald, the parallel orientation of the spin and the magnetic field $B$, leads to accretion of charged particles to the black hole up to values corresponding to a balanced state stopping the accretion, and the black hole charge stays to be \cite{Wald:1974:PHYSR4:} 
\beq
 Q_{\rm W} = 2Ba. \label{WaldCharge}
\eeq

For the sake of simplicity, we will consider in this paper only two limit scenarios:
\begin{itemize}
\item non-charged black hole with $Q=0$,
\item black hole with Wald charge $Q=Q_{\rm W}$.
\end{itemize} 

Note that in the case of non-rotating \Schw{} black hole spacetime with $a=0$, or in the case of the flat spacetime, the electromagnetic four-vector potential $A^{\alpha}$ takes the form 
\beq
A_{t} = 0, \quad A_{\phi} =  B g_{\phi\phi} / 2, \label{AtApFlat}.
\eeq
For the asymptotic limit of rotating Kerr spacetimes ($a\neq0$), the electromagnetic four-vector potential $A^{\alpha}$ takes form 
\beq
A_{t} = - B a , \quad A_{\phi} = B g_{\phi\phi} / 2, \label{AtApKerr}
\eeq
for both $Q=0$ and $Q=Q_{\rm W}$ scenarios. Since the asymptotic limit of the Kerr metric is represented by the flat metric (\ref{KerrFlat}), the existence of the non-zero but constant component $A_{t}$ represents a technical effect, and it leads to a re-definition of the energy level at infinity only, as we will see later.


\subsection{Hamiltonian formalism and reduction to the two-dimensional dynamics \label{reduction2D}}


Hamiltonian of dynamics of a charged test particle in the combined gravo-magnetic field can be written in the form \cite{Mis-Tho-Whe:1973:Gra:,Wald:1984:book:} 
\beq
  H =  \frac{1}{2} g^{\alpha\beta} (\cp_\alpha - q A_\alpha)(\cp_\beta - q A_\beta) + \frac{1}{2} \, m^2
  \label{particleHAM},
\eeq
where the kinematic four-momentum $\p^\mu = m u^\mu$ of a test particle with the mass $m$ and charge $q$ is related to the generalized (canonical) four-momentum $\cp^\mu$ by the relation
\beq
 \cp^\mu = \p^\mu + q A^\mu. \label{particleMOM}
\eeq

The Hamilton equations read 
\beq
 \frac{\d x^\mu}{\d \af} \equiv \p^\mu = \frac{\partial H}{\partial \cp_\mu}, \quad
 \frac{\d \cp_\mu}{\d \af} = - \frac{\partial H}{\partial x^\mu}, \label{Ham_eq}
\eeq
where we introduced the dimensionless affine parameter $\af$ related to the particle proper time $\tau$ by the relation $\af=\tau/m$.

Due to the symmetries of the background spacetime (\ref{KerrMetric}) and the related uniform configuration of the magnetic field (\ref{emAt}-\ref{emAp}), one can easily find the existing conserved quantities related to the particle motion, which are the (covariant) energy and axial angular momentum 
\bea
 -E &=& \cp_t = g_{tt} p^t + g_{t\phi} p^{\phi} + q A_{t}, \label{symE} \\
  L &=& \cp_\phi = g_{\phi\phi} p^\phi + g_{\phi t} p^{t} + q A_{\phi} . \label{symL}
\eea

Introducing for convenience the specific energy, specific axial angular momentum, and specific intensity of the electromagnetic interaction by the relations 
\beq
{\cal{E}} = \frac{E}{m}, \quad {\cal{L}} = \frac{L}{m}, \quad {\cb} = \frac{q B}{2m}, \label{convenience}
\eeq
one can rewrite the Hamiltonian (\ref{particleHAM}) in the explicit form
\beq 
H = \frac{1}{2} g^{rr} \p_r^2 + \frac{1}{2} g^{\theta\theta} \p_\theta^2 + \HP(r,\theta), \label{HamHam} 
\eeq
where the potential part of the Hamiltonian $\HP$ for the test particle with specific charge $\tilde{q} = q/m$ reads 
\bea 
2 \HP &=&  g^{tt} (\ce + \tilde{q} A_{t})^2 - 2 g^{t\phi}(\ce +\tilde{q}A_{t}) (\cl-\tilde{q}A_{\phi}) \nonumber \\
			&& + g^{\phi\phi} (\cl-\tilde{q}A_{\phi})^2 + 1.	\label{HamP}
\eea

Let just note that the parametrization (\ref{convenience}) implies re-definition of the equations related to the asymptotically uniform magnetic field (\ref{WaldCharge}) and (\ref{AtApKerr})
\beq
\tilde{q}Q_{\rm W} = 4 a \cb, \quad \tilde{q}A_{t} = -2 a \cb, \quad \tilde{q}A_{\phi} = g_{\phi\phi} \cb.
\eeq

The Hamiltonian (\ref{HamHam}) has now only two degrees of freedom related to the coordinates $r$ and $\theta$, the phase space is four dimensional $(r,p_r,\theta,p_\theta)$. The dynamics of the system can be effectively described by the $r=r(\tau)$ and $\theta=\theta(\tau)$ evolution relations that can be limited by the boundaries of the motion governed by an effective potential that is a function of the radial and latitudinal coordinates only. We can get rid of the coordinates $t$ and $\phi$ and the knowledge of their evolution laws, $t=t(\tau)$ and $\phi=\phi(\tau)$, is redundant in the context of our study. 
Of course when plotting complete 3D trajectory the knowledge of the evolution law of the $\phi$-coordinate will be necessary, and for the two-particle collisions at a given time event, also the $t$-coordinate evolution will be needed -- such evolution relations can be easily obtained from (\ref{symE}-\ref{symL}). 

\begin{figure*}
\includegraphics[width=\hsize]{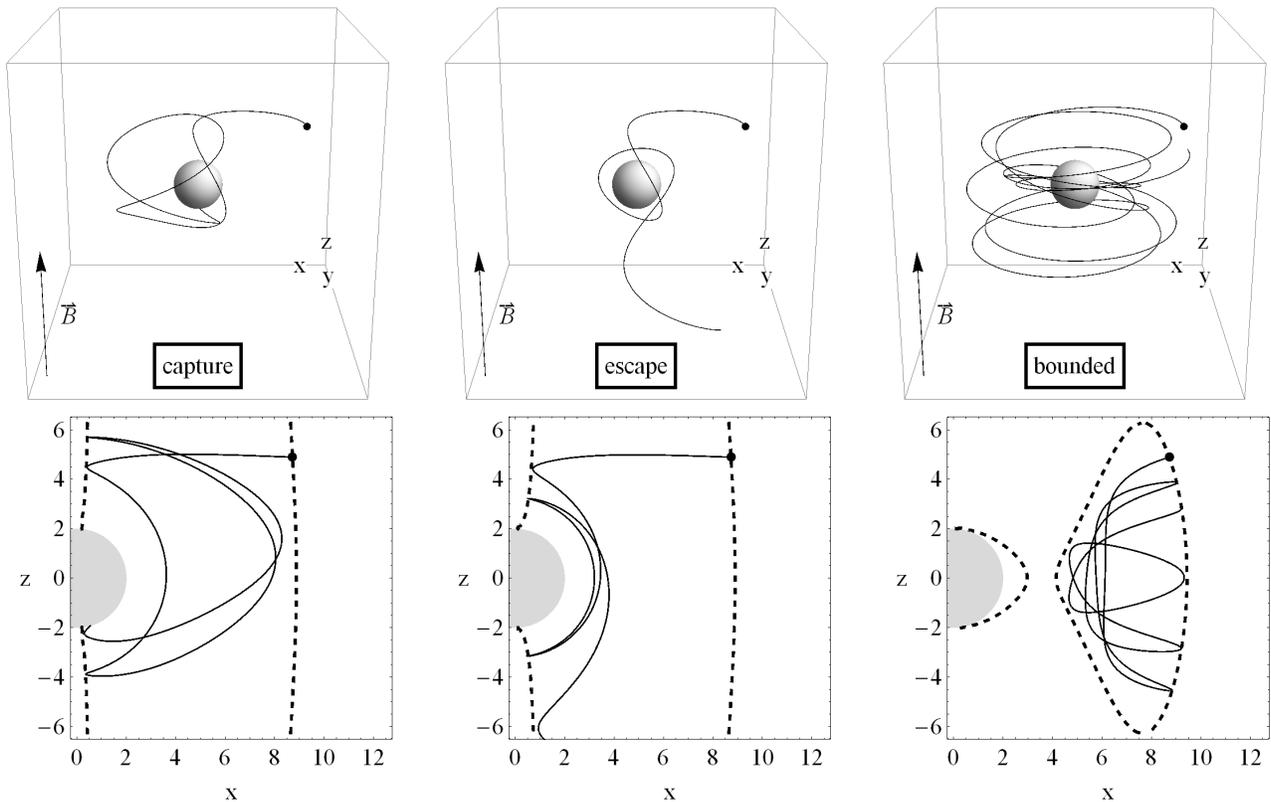}
\caption{
\label{orbitTT}
Typical orbits of charged particles in neighbourhood of a black hole immersed in an uniform magnetic field. In all there cases the charged particle is starting from the point $r_0=8.5,\theta_0\doteq~1.06$ but they differ in angular momentum $\cl$ (i.e. $u^\phi$), and energy $\ce$. We used $\cl=4, \ce\doteq~8.3$ (capture), $\cl=7, \ce\doteq~8.6$ (escape) and $\cl=68, \ce\doteq~14.8$ (bounded). In the upper row 3D trajectories are given, in the lower row 2D ($y=0$) sections of the trajectories and the effective potential are represented.
}
\end{figure*}

The presented reduction to the two-dimensional dynamics $r,\theta$ will help us to clearly distinguish qualitatively what is going on in the charged particle dynamics. In Fig. \ref{orbitTT} we plotted the complete 3D charged particle orbits and also the equivalent 2D reduced charged particle trajectories. In the reduced 2D trajectories we can clearly distinguish the role of the energetic boundary for the motion and how the particle trajectories bounce from the boundaries.

The charged test particle motion is limited by the energetic boundaries implied by the condition $\HP=0$ (\ref{HamP}) and given by the relation 
\beq
 \ce = V_{\rm eff}(r,\theta), \label{MotLim}
\eeq
where the effective potential $V_{\rm eff}(r,\theta)$ of the charged test particle motion takes the form \cite{Kop-etal:2010:APJ:} 
\beq
 V_{\rm eff} (r,\theta) = \frac{ -\beta + \sqrt{\beta^2 - 4\alpha\gamma} }{ 2 \alpha }, \label{VeffCharged}
\eeq
where we have used the abbreviations 
\bea
 \alpha &=& -g^{tt}, \\
 \beta  &=& 2 [ g^{t\phi} (\cl-\tilde{q}A_{\phi}) - g^{tt}\tilde{q}A_{t} ], \\
 \gamma &=& -g^{\phi\phi}(\cl-\tilde{q}A_{\phi})^2 - g^{tt} \tilde{q}^2 A_{t}^2 \nonumber \\
				&&	+ 2 g^{t\phi}\tilde{q} A_{t} (\cl-\tilde{q}A_{\phi}) -1 .
\eea
Discussing the features of the effective potential (\ref{VeffCharged}), we can determine some basic properties of the charged particle dynamics without solving the equations of motion, namely, we can determine the boundaries of the motion. 

Properties of the effective potential $V_{\rm eff}(r,\theta;a,\cl,\cb)$ (\ref{VeffCharged}) related to the motion in the combined field of Kerr black holes and the asymptotically uniform magnetic field were already explored in \cite{Kov-etal:2010:CLAQG:}, and in the simpler case of \Schw{} black holes in \cite{Kol-Stu-Tur:2015:CLAQG:}. Since the general form of the effective potential $V_{\rm eff}(r,\theta;a,\cl,\cb)$ is quite complex, we discuss here only the properties relevant for the purposes of the present paper. 

The $V_{\rm eff}(r,\theta;a,\cl,\cb)$ will be considered here only outside the event horizon of the Kerr black hole. The motion of a charged test particle can be classified according two criteria. We can distinguish the particle motion in relation to the rotation of the Kerr black hole or we can relate the angular momentum of the particle and the direction of the magnetic field. The motion can be thus  
\begin{itemize}
\item {\it prograde}, $a \cl > 0$ - particle is orbiting the black hole in the same direction as black hole rotation
\item {\it retrograde}, $a \cl < 0$ - particle is orbiting the black hole with opposite orientation as the black hole rotation. 
\end{itemize}
In the present paper we focus mainly on the prograde $a \cl > 0$ type of motion, since this kind of the charged particle motion could be more relevant in realistic accretion processes. In the \Schw{} black hole spacetime the prograde and retrograde types of the motion coincide.
In relation to the magnetic field we can distinguish again two main classes of the motion: 
\begin{itemize}
\item[-] {\it minus configuration} $\cb\cl<0$ - the magnetic field and angular momentum parameters have  opposite signs 
\item[+] {\it plus configuration} $\cb\cl>0$ - the magnetic field and angular momentum parameters have identical signs.  
\end{itemize}
The effective potential $V_{\rm eff}(r,\theta;a,\cl,\cb)$ clearly demonstrates the symmetry $(a,\cb,\cl)$ $\leftrightarrow(-a,-\cb,-\cl)$ and the combinations of the prograde and retrograde motion with the minus and plus configurations govern in principle four independent types of the effective potential $V_{\rm eff}(r,\theta;a,\cl,\cb)$ behaviour. Hereafter we will consider $a>0$ and only prograde trajectories with positive angular momentum $\cl>0$ -- particle will rotate counter-clockwise. For covering both minus and plus magnetic field configurations we will be using negative $\cb<0$ and positive $\cb>0$ values of the parameter $\cb$. For details see \cite{Kol-Stu-Tur:2015:CLAQG:}. 

The stationary points of the effective potential $V_{\rm eff}(r,\theta;a,\cl,\cb)$, governing circular orbits of the charged test particles, are determined by the conditions 
\beq
  \partial_r V_{\rm eff}(r,\theta) = 0, \quad \partial_\theta V_{\rm eff}(r,\theta) = 0.  \label{extrmy}
\eeq
For the \Schw{} black holes all the local extrema of the effective potential $V_{\rm eff}(r,\theta;a,\cl,\cb)$ are located in the equatorial plane $\theta=\pi/2$ only, but for the rotating Kerr black holes also the off-equatorial extrema giving circular orbits are possible \cite{Kov-etal:2010:CLAQG:}. 

The motion of charged test particles governed by the effective potential $V_{\rm eff}(r,\theta;a,\cl,\cb)$ is generally chaotic, but there exist regions where a regular motion is allowed, for example in the vicinity of the stable equilibrium points corresponding to the minima of the effective potential $V_{\rm eff}(r,\theta;a,\cl,\cb)$. 

In Fig. \ref{orbitTT} we plotted characteristic trajectories of the charged test particles, representing namely the capture by the black hole, escape along the symmetry axis to infinity, and bounded motion in the strong gravity of the black hole. In the following we focus our attention to the case of the charged particles that can escape to infinity.


\subsection{Escape to infinity along the magnetic field lines}


The charged particle motion in the effective potential $V_{\rm eff}(x,z;a,\cl,\cb)$ (\ref{VeffCharged}) is always bounded in the $x$-direction due to the influence of the magnetic field, $\sim~\cb^2~x$, and the angular momentum, $\sim~\cl/x$, terms, but the energetic boundary for the motion $\ce=V_{\rm eff}(x,z)$ (\ref{MotLim}) can be open in the $z$-direction, enabling the charged particles to escape to infinity along the $z$-axis.

In the Kerr black hole spacetimes, the effective potential $V_{\rm eff}(x,z;a,\cl,\cb)$ of the charged particle dynamics takes in the asymptotic limit $z\rightarrow\infty$ the form  
\beq
 V_{\rm eff}(x,z\rightarrow\infty)  = 2 a \cb + \sqrt{1+ \left( \frac{\cl}{x} - \cb x \right)^2}. \label{VeffZ}
\eeq 
This potential, as a function of $x$ only, has local minima located at
\beq
 x_{\rm min}^2 = \cl / \cb,
\eeq
where the corresponding energy (\ref{MotLim}) arrives to the form 
\beq
 \ce_{\rm min} = 
\Big\{ 
\begin{array}{l @{\quad} c @{\quad} l} 
2 a \cb + 1 & \textrm{for} & \cb \geq 0 \\ 
2 a \cb + \sqrt{1 - 4 \cb \cl} & \textrm{for} & \cb < 0 \\ 
\end{array} \label{minE}
\eeq 
The charged test particles can escape to infinity along the $z$-axis if their energy is high enough, namely, 
\beq
 \ce \geq \ce_{\rm min}. \label{escapeCON}
\eeq

Expressing the flat Minkowski metric in the cylindrical coordinates $(t,\rho,\phi,z)$, the effective potential of the charged test particle motion in the flat spacetime with the uniform magnetic field takes the form 
\beq
 V_{\rm eff(flat)}(\rho)  = \sqrt{1+ \left( \frac{\cl}{\rho} - \cb \rho \right)^2}. \label{VeffFLAT}
\eeq 
If we compare the effective potential (\ref{VeffFLAT}) with the expression (\ref{VeffZ}), we can from (\ref{MotLim}) clearly see that the energy of charged test particles in the Kerr spacetime, $\ce$, cannot be interpreted as energy measured at infinity where the uniform magnetic field exists. The energy measured at infinity should be in this case given by the relation 
\beq
 \ce_\infty = \ce - 2 a \cb. \label{Einfty}
\eeq 
The difference between the $\ce_\infty$ and $\ce$ is caused by existence of a non-zero but constant $t$-component of the electromagnetic potential $A_{t}\neq0$ related to the magnetic field. 


\begin{figure*}
\includegraphics[width=\hsize]{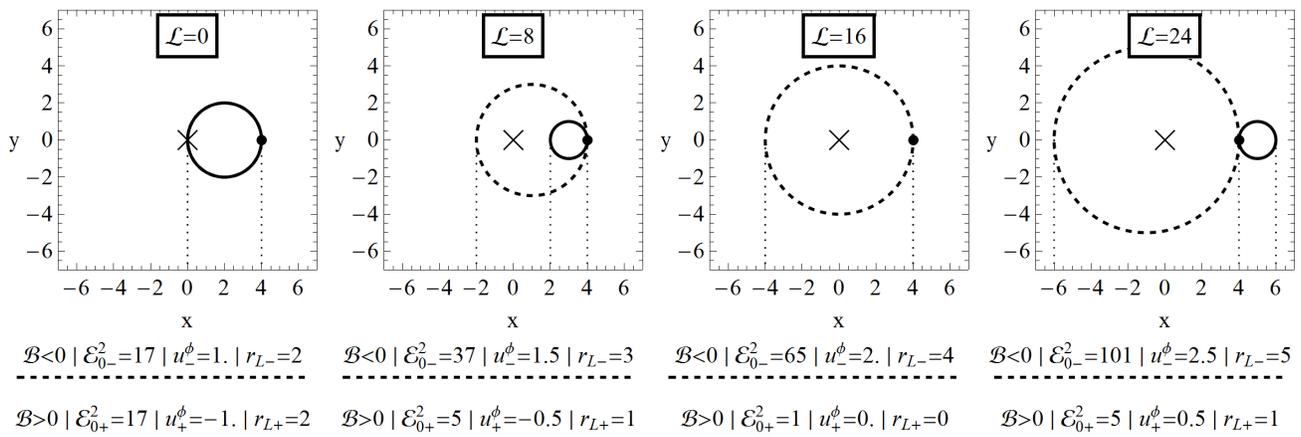}
\caption{ \label{flatORB}
Influence of the angular momentum parameter $\cl$ on the charged particle orbit in the uniform magnetic field and flat spacetime. 
The motion in the $z$-direction is decoupled from complete dynamics and the charged particle trajectory is projected into a circle in the $x-y$ plane, with the Larmour radius $r_{\rm L}$. The particle is orbiting the circle counter-clockwise for $\cb=-1$ and clockwise for $\cb=1$. Due to the symmetry of the dynamics, we consider $\cl\ge~0$ only. (For $\cl\le 0$ the position of circles is symmetric about the centre $\rho = 0$.)
In each figure both the minus, $\cb=-1$ (dashed circle), and the plus, $\cb=1$ (solid circle), configurations are considered for a particle starting from the point with $rho_0=4$, with various energy $\ce_0$ and angular velocity $u^\phi$, and in dependence on the angular momentum parameter $\cl$. 
For $\cl=0$ value both the minus and plus configurations of the parameter $\cb$ share the same orbit. The inner and outer radius of oscillations (\ref{Turn}) is $x_{\rm i,o}\in\{0,\pm4\}$.
For $\cl=8$, the circle radius is bigger for $\cb=-1$ than the circle radius for $\cb=1$. The inner and outer radii of the oscillatory motion along the circle, governed (\ref{Turn}), read $x_{\rm i,o}\in\{\pm2,\pm4\}$. 
For $\cl=16$, the charged particle stops its circular motion in the $\cb=1$ case at $x_i=x_o=\pm4$, while the the inner and outer radii of the oscillatory motion, given by (\ref{Turn}), read $x_{\rm i,o}\in\{\pm4\}$.
For $\cl=24$, the inner and outer radii of the oscillatory motion, given by (\ref{Turn}), read $x_{\rm i,o}\in\{\pm4,\pm6\}$.
}
\end{figure*}

\begin{figure}
\includegraphics[width=0.85\hsize]{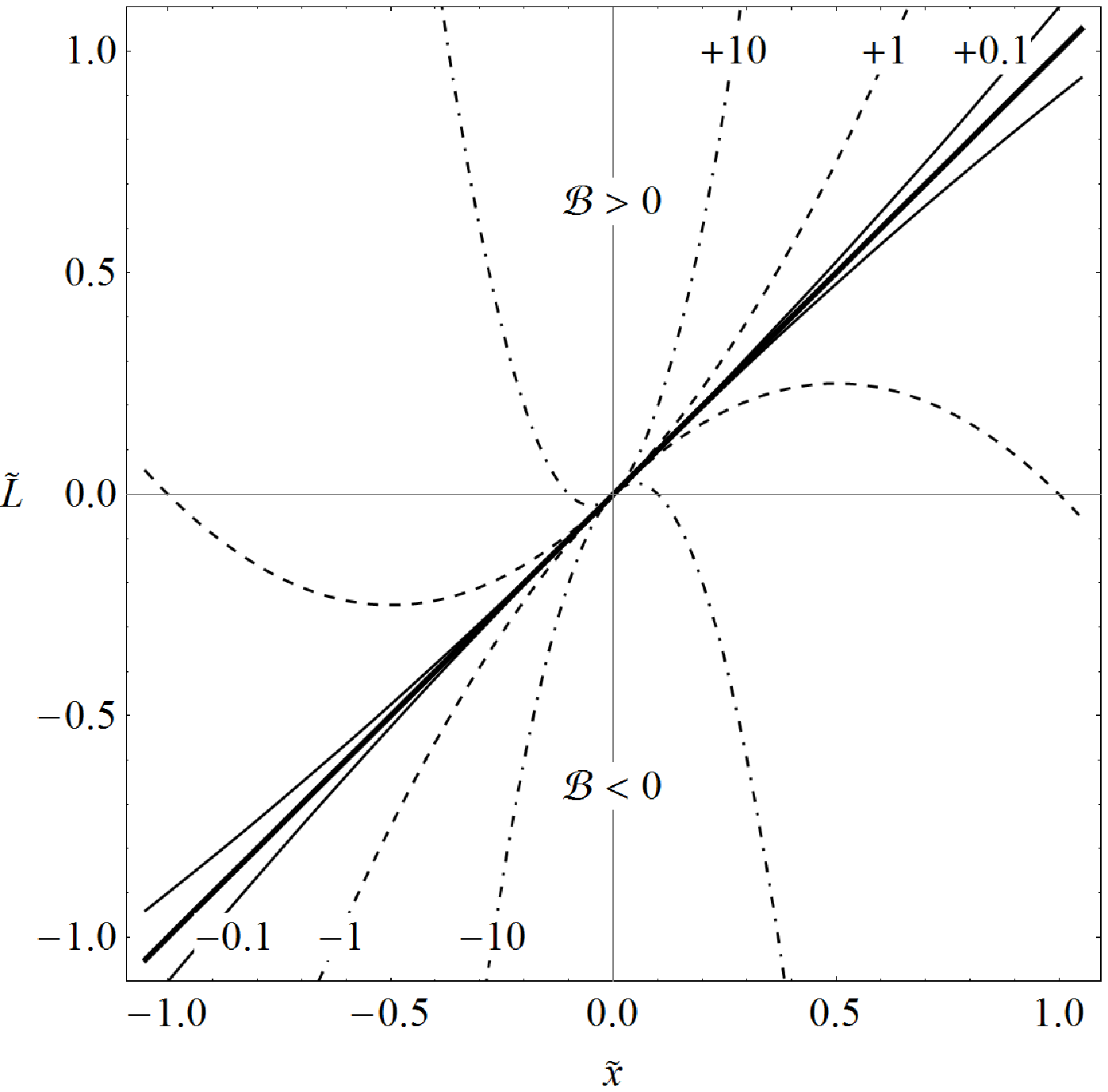}
\caption{ \label{TurnreduFIG}
Position of the inner and outer oscillation radii $\tilde{x}$ in dependence on the $\tilde{\cl}$ and $\cb$ parameters. Simple parabolic dependence of $\tilde{\cl}(\tilde{x},\cb)$ on transformed coordinate $\tilde{x}$ for significant values of parameter $\cb\in{0,\pm0.1,\pm1,\pm10}$. The $\cb=0$ line (thick) is separating sections for $\cb > 0$ (above) and $\cb < 0$ (below).   
}
\end{figure}


\subsection{Charged particle dynamics in asymptotically flat region of the rotating Kerr spacetime}


The charged particles can reach infinity along the $z$-axis, if their energy is bigger then the minimal energy at infinity, see (\ref{escapeCON}). We consider the charged particle dynamics in asymptotically flat limit of the rotating Kerr spacetime (\ref{KerrCyli}) expressed in the cylindrical coordinates $(t,\rho,\phi,z)$ and with external uniform magnetic field given by (\ref{AtApKerr}). The condition $H=0$ giving the energy boundary function related to the Hamiltonian (\ref{particleHAM}) can be written in the form 
\bea
(\ce-2a\cb)^2 = \ce_{\infty}^2 &=& \dot{z}^2 + \dot{\rho}^2 + g_{\phi\phi} \dot{\phi}^2 \nn\\
&=& \dot{z}^2 + \dot{\rho}^2 + \left( \cl/\rho - \cb \rho \right)^2 + 1 \nn\\
&=& \ce^2_{\rm z} + \ce^2_{\rm 0} \label{ENflat}
\eea
where the dot denotes derivative with respect to the proper time, ie., the proper velocity $u^\alpha$
\beq
 u^\alpha = \frac{\d x^\alpha}{\d \tau} = \dot{x}^\alpha. \label{speedU}
\eeq

We define the translational kinetic energy mode in the $z$-direction, $\ce_{z}$, and the 'perpendicular' energy mode, $\ce_{0}$, using the flat space version of the Hamiltonian and the decomposition of the total energy given by Eq.(\ref{ENflat}) (Note that sometimes the translational mode is denoted as longitudal, while the perpendicular mode is denoted as transverse.)
\beq
 \ce^2_{\rm z} = \dot{z}^2, \quad \ce^2_{\rm 0} = \dot{\rho}^2 + \left( \cl/\rho - \cb \rho \right)^2 + 1.
\eeq
The 'perpendicular' energy $\ce_0$ is composed from the rest mass term ($1$-term), kinetic energy in the radial $\rho$-direction ($\dot{\rho}$ term) and the kinetic energy in the $\phi$-direction ($\rho^2\dot{\phi}$ term). The last term can be easily rewritten using the conserved quantity $\cl$, angular velocity $u^\phi=\dot{\phi}$ and the $\phi$-component of the electromagnetic four potential $A_{\phi}$ 
\beq
 \cl = g_{\phi\phi} u^\phi + \tilde{q} A_\phi = \rho^2 ( u^\phi + \cb ). \label{clflat}
\eeq

We can distinguish four different energies: energy of the charged particle $\ce$ - constant of the motion, energy resumed at infinity $\ce_\infty$ (\ref{Einfty}) - also constant of the motion, the kinetic energy in the $z$-direction $\ce_{z}$, and the 'perpendicular' energy $\ce_{0}$. They are related by  
\beq
 (\ce-2a\cb)^2 = \ce_{\infty}^2 = \ce^2_{\rm z} + \ce^2_{\rm 0} . 
\eeq

The dynamics of charged particles in the flat spacetime with uniform magnetic field is regular and can be solved analytically. The motion in the $z$-direction, along the magnetic field lines, can be separated from the dynamics in the $x-y$ plane, see Eq. (\ref{ENflat}). Dynamics in the $z$-direction corresponds to the translational (linear) motion with constant velocity $\dot{z}$, while in the $x-y$ plane an uniform circular motion occurs with the Larmor radius $r_{\rm L}$ and the Larmor period $T_{\rm L}$ given by the relations 
\beq
 r_{\rm L} = \frac{\rho_0 u^\phi}{2\cb}, \quad T_{\rm L} = \frac{\pi}{\cb},
\eeq
where $\rho_0$ denotes the particle initial position (its radial coordinate), and $u^\phi$ is charged particle angular velocity (\ref{clflat}) given by the relation 
\beq
 u^\phi = \cl / \rho_0^2 - \cb. \label{uphi}
\eeq
Typical examples of the charged particle circular orbits in the $x-y$ plane are plotted in Fig. \ref{flatORB} for significant values of the canonical angular momentum $\cl$, and the corresponding values of energy $\ce_0$, the Larmor radius $r_{\rm L}$, and the angular velocity $u^\phi$. 

The Lorentz force acting on charged particles moving in the uniform magnetic field directed along the $z$-axis is generated by the velocity in the $x-y$ plane only. As long as the charged particles are moving only along the magnetic field lines in the $z$-direction, with $u^\phi=0$, the charged particle motion is not influenced by the magnetic field. The condition $u^\phi=0$ can be fulfilled only by the plus configurations with $\cl \cb>0$. 

The charged particle motion in the flat spacetime with the uniform magnetic field is a combination of the uniform circular motion in the $x-y$ plane and the linear translational motion along the $z$-axis, giving a helical motion as a complete 3D trajectory. In order to reflect the position of the turning points of the oscillatory motion relative to the symmetry axis defined by $\rho = 0$, or equivalently, $x=0, y=0$, we use the restriction (projection) of the motion to the plane $y=0$. Using the 2D reduction, separating the $\phi$-coordinate motion (see section \ref{reduction2D}), we can interpret the components of the 'perpendicular' energy $\ce_0$ as the $\rho=x$-coordinate kinetic energy ($\dot{\rho}$ term) and 'rest' energy of the particle (the rest 1-term). The charged particle is bounded in its radial ($\rho=x$) motion, by the angular momentum $\cl$ and the magnetic field $\cb$ barriers, so the charged particle will follow an oscillatory motion in the $\rho=x$-direction when we can consider both positive and negative values of the coordinate $x$. The inner and outer radius of the oscillatory motion in the $x$-direction for the motion with $\ce > 1$ are given by the formula
\beq
 x_{\rm o,i} = \frac{ -\sqrt{\ce_{0}^2-1} \pm \sqrt{\ce_{0}^2-1 + 4 \cb \cl} }{2 \cb}. \label{Turn}
\eeq
To every positive value of $x_{\rm o,i}$ there exist negative value as well. Clearly, in the case of $\cl\cb < 0$, the axial angular momentum has to be limited by the relation $\cl>(1-\ce_0^2)/4\cb$.

The turning points in the $x$-coordinates, $x_{\rm o,i}$, can be determined by the condition  
\beq
 \tilde{\cl}(\tilde{x},\cb) = \tilde{x} (\cb \tilde{x} +1),  \label{Turnredu}
\eeq
where we introduce transformed coordinate $\tilde{x}$ and transformed angular momentum $\tilde{\cl}$ by the relations 
\beq
 \quad \tilde{x} = \frac{x}{\sqrt{\ce_0^2-1}},\quad \tilde{\cl} = \frac{\cl}{\ce_0^2-1}.
\eeq
The simple parabolic dependence of $\tilde{\cl}(\tilde{x},\cb)$ on transformed coordinate $\tilde{x}$ for significant values of parameter $\cb$ is illustrated in Fig. \ref{TurnreduFIG}. 

We can express the 'perpendicular' energy $\ce_0$ using the coordinates of the turning points of the oscillatory motion in the form 
\beq
 \ce_0^2 = \cb^2 (x_{\rm i} + x_{\rm o})^2 +1 \label{E0xixo}
\eeq

The oscillatory motion in the $\rho=x$-direction vanishes ($\dot{\rho}=0$) when the inner and outer radii coincide at 
\beq
 \rho_{\rm i} = \rho_{\rm o} = \rho_{\rm min} = \sqrt{| \cl / \cb |}, \label{Xmin}
\eeq
and the 'perpendicular' energy $\ce_0$ reaches its minimal value of 
\beq
\ce_{\rm 0(min)} = 
\Big\{ 
\begin{array}{l @{\quad} c @{\quad} l} 
1 & \textrm{for} & \cl\cb \geq 0 \\ 
\sqrt{1 - 4 \cb \cl} & \textrm{for} & \cl\cb < 0 . \\ 
\end{array}
\label{E0min}
\eeq 
Notice that $\ce_{\rm 0(min)}=\ce_{\rm min}-2a\cb$ due to Eq.(\ref{minE}). The role of the orientation of the magnetic field relative to the angular momentum of the particle in the position of the turning points of the oscillatory motion is demonstrated in Fig.\ref{flatORB}. 


\begin{figure*}
\includegraphics[width=\hsize]{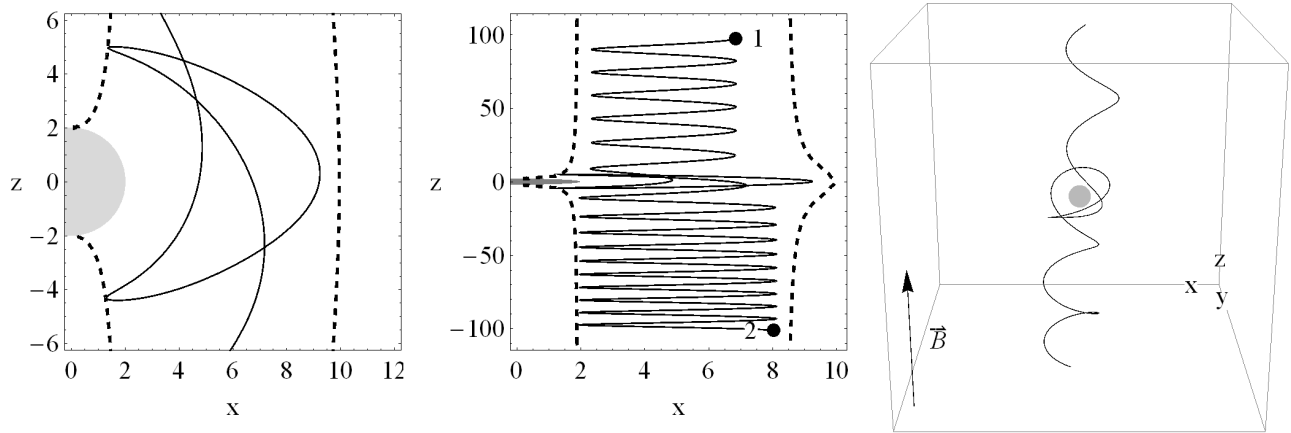}
\caption{ \label{trans}
Transmutation of charged particle trajectory. Charged particle is starting at almost flat spacetime region - far away from black hole, moving towards the black hole along the magnetic field lines, which are parallel to the $z$-axis. The particle is crossing the equatorial plane near the black hole horizon, where effect of gravitational and magnetic fields is strong, the particle dynamics si chaotic, and the extension of the charged particle radial oscillations is changed. 
The presented charged particle trajectory (black curve), with angular momenta $\cl=16$ and energy $\ce\doteq10$, can start from the points (1) $(x_0,y_0,z_0)=(6.8, 0, 97.7)$ or (2) $(x_0,y_0,z_0)=(-5.3, -6, -101)$.
The three sub-figures are representing the same particle trajectory, but for different views. The \Schw{} black hole is non-rotating, $a=0$, the external uniform magnetic field parameter is $\cb=-1$.
}
\end{figure*}


\section{Black holes in the asymptotically uniform magnetic field as charged particle accelerator}


\subsection{Transmutation of energy modes of charged particle motion}


The effective potential $V_{\rm eff}(x,z)$ for the charged test particle motion can be open towards infinity along the $z$-axis, allowing some charged particles to escape to infinity - such kind of motion along a corridor concentrated about the axis of symmetry of the black hole spacetime is not allowed for neutral test particles. This type of motion provides an interesting astrophysical implication as it can be used for modelling of relativistic jets, if the velocity of particles escaping along the axis of rotation can be relativistic. 

Let us consider a charged particle far away from the black hole, where the spacetime can be considered as flat, orbiting around a $z$-axis parallel to the symmetry axis of the Kerr spacetime with some 'perpendicular' energy $\ce_0$ and moving with some kinetic energy $\ce_{\rm z}$ in the $z$-direction towards the Kerr black hole, see point (1) at Fig. \ref{trans}. The motion is quite regular as far as the charged particle is far away from the black hole. Close to the black hole horizon, the particle motion becomes chaotic and after a scatter on the combined gravitational and magnetic field represented by the gravo-magnetic potential given by Eq. (\ref{VeffCharged}), the charged particle escapes towards infinity. Once again the motion is regular as the particle reaches the flat spacetime region, see point (2) at Fig. \ref{trans}. 

However, there is an important change in the particle dynamics as points (1) and (2) are compared -- the oscillation amplitude (inner, $x_{\rm i}$, and outer, $x_{\rm o}$, radii) in the $\rho$($x$)-direction has been changed, moreover, also the $z$-axis of the final motion could be shifted in comparison to the axis of the initial phase of the motion. This means that the 'perpendicular' energy $\ce_0$ between the points (1) and (2) has changed -- $\ce_{0(1)}\neq\ce_{0(2)}$, see Eq. (\ref{E0xixo}). Since the charged test particle energy at infinity $\ce_{\infty}$ (\ref{ENflat}) has to be constant during the motion, we obtain also a change in the kinetic energy mode along the $z$-axis -- $\ce_{\rm z(1)}\neq\ce_{\rm z(2)}$. The particle speed along the $z$-axis has been changed by scattering in the effective potential close to the black hole horizon. We observe a transmutation effect -- energy transmission between the energy modes $\ce_{z}$ and $\ce_{0}$.

We see an increase in the amplitude of oscillations in the radial $\rho$-direction, so there has to be a decrease in the particle proper velocity (\ref{speedU}) along the $z$-axis, $ u^z_{(1)}>u^z_{(2)}$ -- the charged particle has been {\it slowed down} in the effective potential. Since the equations of motion (\ref{Ham_eq}) are independent of the time orientation, the charged particle can also start in the point (2) and go to the point (1) and hence a {\it speed-up} in the effective potential could also occur.

Clearly, the energies $\ce$, $\ce_\infty$, $\ce_{\rm z}$ and $\ce_0$ are constants of the charged particle motion in the flat spacetime, therefore, no transfer between $\ce_{\rm z}$ and $\ce_0$ energy modes (transmutation of the energy modes) is possible there. On the other hand, in vicinity of black holes, only the total covariant energy $\ce$ is conserved, the energy modes $\ce_{\rm z}, \ce_0$ can be changed during the scatter in the strong gravo-magnetic field and the transmutation effect can work. 
The effect of energy transmission between the energy modes $\ce_{\rm z}$ and $\ce_0$ implies a corresponding change of the charged particle speed along the $z$-axis. All the kinetic energy mode $\ce_{\rm z}$ can be transmitted to the 'perpendicular' energy mode $\ce_0$ and the charged particle will just stop its motion along the $z$-axis, while the oscillations in the $\rho$-direction increase to the maximal limit. On the other hand, whole the 'perpendicular' energy mode $\ce_0$ cannot be transmitted to the kinetic energy mode $\ce_{\rm z}$ -- there always remains some inconvertible energy in the $\ce_0$ energy mode determined by the minimum energy $\ce_{\rm 0(min)}$ -- see (\ref{E0min}). 

The presented energy transmutation effect, i.e., the interchange $\ce_{\rm z}\leftrightarrow\ce_0$, does not require the black hole rotation (and phenomena related to the ergosphere as in the Penrose process, or Blandford-Znajek processes) and works even in the \Schw{} spacetime. For the energy transmutation effect of energy modes of the charged particle motion, there is no energy mining from the black hole, the effect is purely 'mechanical' - just the energy modes interchange their energy; the nature of this effect lies in the chaotic nature of the charged particle motion.

\begin{figure*}
\includegraphics[width=\hsize]{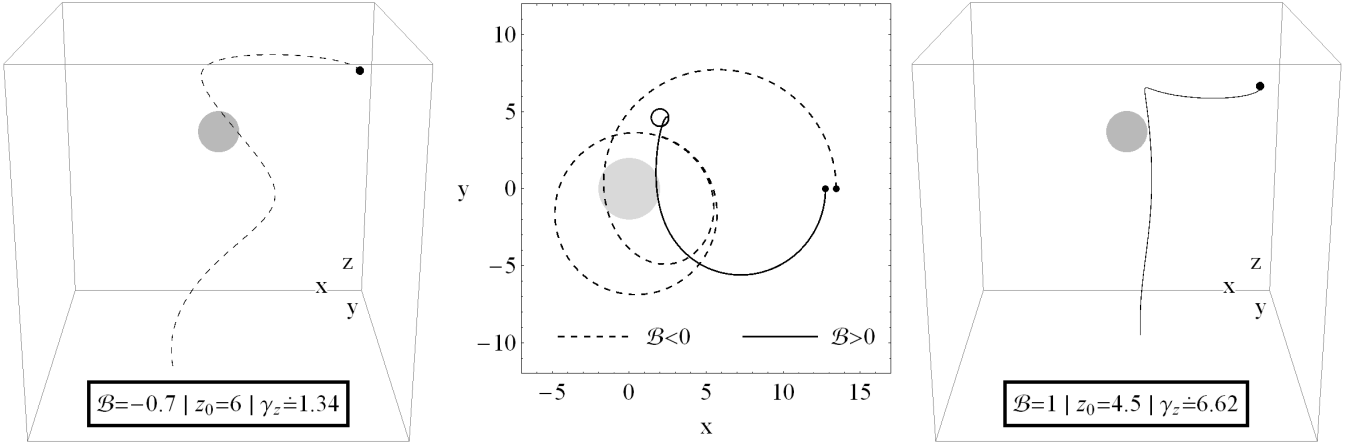}
\caption{ \label{largeGZ}
Trajectories with large escape velocities and gamma factor $\gamma_{\rm z}$ given by chaotic scattering process, see Fig \ref{chaotic}. The trajectories with largest $\gamma_{\rm z}$ are moving almost along straight trajectories for positive $\cl\cb>0$ configuration (right), while for negative $\cl\cb<0$ configuration (left) some orbital movement around magnetic field lines always remains. We compare both trajectories in the $x-y$ pane projections (middle).
}
\end{figure*}


\subsection{Escape velocities in the chaotic scattering \label{velocitiesSEC}}


The possibility of strong acceleration of the linear translational motion along the magnetic field lines is from the astrophysical point of view one of the most relevant applications of the charged particle dynamics in strong gravo-magnetic fields as it can be applied as a model of acceleration of relativistic jets in Active Galactic Nuclei (AGN) and microquasars. Let us consider a charged particle orbiting a Kerr black hole, following a quasi-circular motion in small distance $z_{\rm 0}$ from the equatorial plane. Then the energy transmutation effect, $\ce_{\rm z}\leftrightarrow\ce_0$, discussed above, can enter the play due to the chaotic nature of the equations of motion in the strong gravitational field of the black hole combined with the asymptotically uniform magnetic field, leading eventually to escape of the charged particle along the $z$-axis with relativistic velocity as measured at infinity. 

In the asymptotically flat spacetime far away from the Kerr black hole, filled by the uniform magnetic field, the charged test particle motion is determined by the coordinate velocity $v^\alpha$, and proper velocity $u^\alpha$, given by the expressions 
\beq
  v^\alpha = \frac{\d x^\alpha}{\d t}, \quad u^\alpha = \frac{\d x^\alpha}{\d \tau} = \gamma v^\alpha, \quad \gamma = \frac{\d t}{\d \tau}, \label{speedDEF}
\eeq
where we have introduced the Lorentz $\gamma$ factor in order to relate the proper and coordinate velocities 
\beq
 \gamma = u^{t} = \frac{\d t}{\d \tau} = \ce + \tilde{q}A_{t} = \ce - 2 a \cb = \ce_\infty. \label{gammaDEF}
\eeq
The individual components of the proper velocity $u^\alpha$ are: velocity of the oscillatory motion in the radial $\rho$ coordinate  - $u^{\rho}$, the ejection velocity along the $z$-axis, i.e. the symmetry axis of the asymptotic motion - $u^{z}$, the orbital velocity of the revolving motion around the $z$-axis - $u^{\phi}$, and the time component of the 4-velocity related to the energy of the motion - $u^{t}=\gamma=\ce_{\infty}$. 

Ejection speeds $u_{\rm z}=u^{\rm z}, v_{\rm z}=v^{\rm z}$ and the corresponding Lorentz $\gamma_{\rm z}$ factor related to the $z$-axis can be expressed using the energetic relations (\ref{ENflat})
\beq
u_{\rm z} = \ce_{\rm z},\quad v_{\rm z} = \frac{\ce_{\rm z}}{\ce_\infty}, \quad \gamma_{\rm z} = \frac{1}{\sqrt{1-v_{\rm z}^2}} =\frac{\ce_\infty}{\ce_0}. \label{speedZ}
\eeq
The velocities and the related $\gamma$ factor can take values $u_{\rm z}\in\langle~0,\infty)$, $v_{\rm z}\in\langle~0,1)$ and $\gamma_{\rm z}\in\langle~1,\infty)$. The first value of allowed interval is valid for vanishing motion along the $z$-axis, the second number occurs when the charged particle is moving along the $z$-axis with the speed of light, representing obviously an inaccessible limit for the charged particle with mass $m\neq0$.

\begin{figure*}
\includegraphics[width=\hsize]{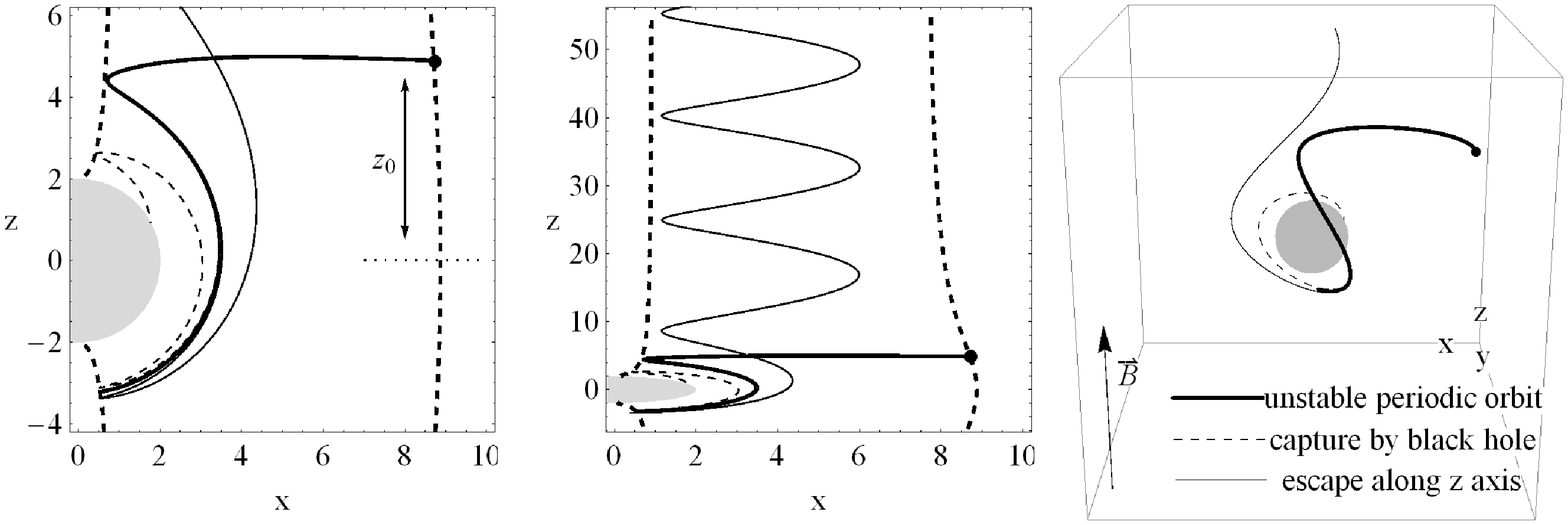}
\caption{ \label{escapeTT}
Unstable periodic orbits separating capture and escape trajectories. All sub-figures represent different views of the same charged particle trajectory around the \Schw{} black hole, $a=0$, immersed into the external uniform magnetic field with parameter $\cb=-1$.
Unstable periodic orbit (thick curve) with initial conditions $r_0=10,\theta_0\doteq1,\cl=7,\ce\doteq8.6$ is trajectory of a particle that returns to the same initial point $r_0,\theta_0$ (periodic), but slightly different initial conditions will produce completely different trajectories (unstable). 
Even small difference in initial conditions can have totally different outputs in the neighbour of unstable periodic orbit; we plotted trajectory captured by the black hole and also trajectory backscattered, escaping to infinity along the $z$ axis.
}
\end{figure*}

The limit on maximal speed $v_{\rm z}$, or gamma factor $\gamma_{\rm z}$, of charged particle moving along the $z$-axis (\ref{speedZ}) can be obtained by taking the minimal value of the perpendicular energy $\ce_0$, given by (\ref{E0min}). Then the limiting, maximal value of the $\gamma$ factor for the velocity in the $z$-direction is given by  
\bea
 \cb>0 &:& \gamma_{\rm z(max)} = \frac{\ce_\infty}{\ce_{\rm 0(min)}} = \ce_\infty, \label{speedZmaxP} \\
 \cb<0 &:& \gamma_{\rm z(max)} = \frac{\ce_\infty}{\sqrt{1-4\cb\cl}} . \label{speedZmaxM} 
\eea
If the charged particle is maximally accelerated (\ref{speedZmaxP}-\ref{speedZmaxM}), then the radial oscillations vanish, $u^{\rm \rho}=0$, and for the orbital speed $u^{\phi}$ we obtain relations 
\bea
 \cb>0 &:& u^{\phi} = 0,  \label{minPspeedP} \\
 \cb<0 &:& u^{\phi} = 2 \cb \cl. \label{minPspeedM}
\eea 

For the plus configurations with $\cb>0$, the acceleration of charged test particles due to the energy transmutation effect, $\ce_{\rm z}\leftrightarrow\ce_0$, works pretty well, the velocity along the $z$-axis, $v_{\rm z}$, can reach almost the speed of light, depending only on the initial energy $\ce$. When the speed along the $z$-axis will be maximal, given by (\ref{speedZmaxP}), the charged particle is moving along a straight line parallel to the symmetry axis of the Kerr spacetime, just along a magnetic field line and there will be no orbital motion since $u^{\phi} = 0$, see Fig. \ref{largeGZ}. 
For the minus configurations with $\cb<0$, the acceleration of charged particles is less efficient. The charged particle  with maximal velocity in the $z$-direction (and minimal perpendicular energy) has always a non-zero orbital velocity $u^{\phi} = 2 \cb \cl$ and hence it cannot have the escape velocity in the $z$-direction as large as for the plus-configurations with $\cb>0$ -- see Fig. \ref{largeGZ}.

\begin{figure*}
\includegraphics[width=\hsize]{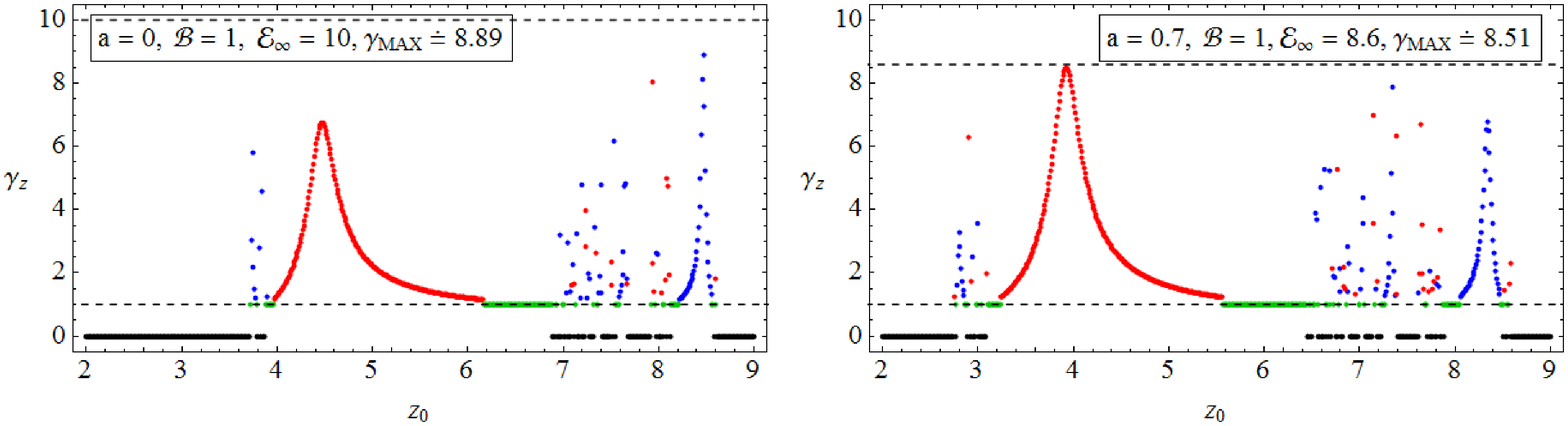}
\includegraphics[width=\hsize]{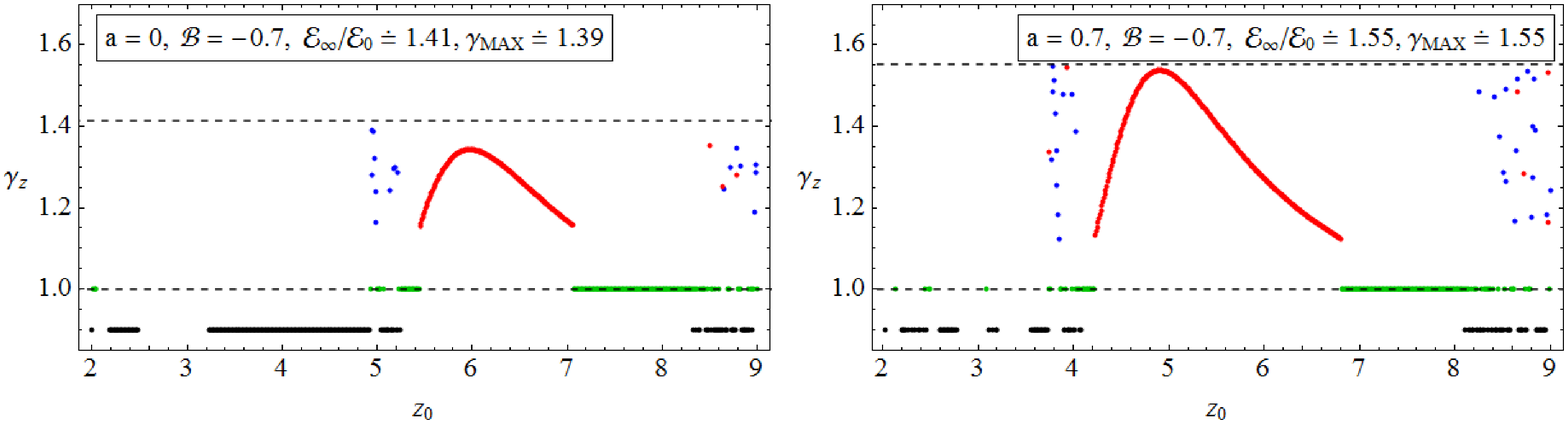}
\caption{ \label{chaotic}
Chaotic scatter of charged particles. The scatter is represented by dependence of charged particle velocity $\gamma_{\rm z}(z_0)$ along the $z$ axis on the initial distance from the equatorial plane, for charged particles scattered in combined gravitational and magnetic effective potential, see Fig. \ref{escapeTT}. All charged particles are starting with energies $\ce=10$ in the neighbourhood of $x_0\sim13$ and $z_0\in(2,9)$ in \Schw{} (left column) or Kerr (right column) black holes spacetime .
The allowed region of $\gamma_{\rm z}$ velocities is bounded by the lower $1$ and upper $E_\infty/E_0$ limits (dashed lines), see eq. (\ref{speedZmaxP}). The trajectories with $\gamma_{\rm z}=1$ are those that not escape to infinity ($r=10^3$) in given time, trajectory with $\gamma_{\rm z}=0$  are captured by the black hole. 
For the positive configuration, $\cb\cl>0$, of magnetic field and angular momentum (first row) we use parameters $\cb=1$ and $\cl=25$, leaving the maximal allowed acceleration to be $E_{\infty}/E_0=10$ for \Schw{} and $E_{\infty}/E_0=8.6$ for the Kerr black hole, with $a=0.7$. For the negative configuration, $\cb\cl<0$, of magnetic field and angular momentum (second row) we use parameters $\cb=-0.7$ and $\cl=17.5$, giving the maximal allowed acceleration to be $E_{\infty}/E_0\doteq~1.41$ for the \Schw{} and $E_{\infty}/E_0\doteq~1.55$ for the Kerr black hole.
Examples of individual escape trajectories with large gamma factors $\gamma_{\rm z}$ can be found in Fig. \ref{largeGZ}.
}
\end{figure*}


We can examine the efficiency of the charged particle acceleration process for the whole set of initial conditions using the chaotic scattering theory as it is presented in chapter 5 of \citep{Ott:1993:book:} and chapter 8 of \citep{Tel-Gru:1993:book:}.

In the classical theory of chaotic scattering we consider a particle with impact parameter $b$ entering a force field represented by an effective potential (scattering region), and characterize result of the scattering process by  the so called scattering angle $\alpha$ of the particle escaping the scattering region. We can define scattering angle (scattering function) $\alpha(b)$ as a function depending on impact parameter $b$, and, of course, on the character of the force field influencing the particle that is reflected by the character of the scattering function. Chaotic scattering theory is dealing with properties of the scattering function $\alpha(b)$, especially when $\alpha(b)$ shows some ''strange'' chaotic behaviour. 
In our problem of a charged particles moving in the gravito-magnetic field, the particle can escape the system only in the $z$-direction corresponding to the field lines of the magnetic field, or it can be captured by the black hole. We will shoot charged particles from some position $z_{\rm s}$ giving initial distance from the equatorial plane (playing the role of the impact parameter), towards to the black hole (scattering region) and we will calculate the escape speeds in the $z$-direction (scattering function) given by the final Lorentz gamma factor $\gamma_{\rm z}$ . 

When the effective potential $V_{\rm eff}(r,\theta)$ of the gravo-magnetic field is for a given energy $\ce$ open towards the black hole horizon and also towards infinity in the $z$-direction, then three quantitatively different types of orbits exist -- captured orbits that cross the black hole horizon, scattered and backscattered orbits that reach $z\rightarrow\pm\infty$, and special family of unstable periodic orbits -- see Fig. \ref{escapeTT}.

The extremely accelerated trajectories have the final $x$-coordinate ($\rho$-coordinate) almost constant and very close to $x_{\rm min}$, given by (\ref{Xmin}). For an efficient energy transmission, $\ce_{\rm z}\leftrightarrow\ce_0$, and hence large acceleration in the $z$-direction, the charged particle necessarily enters the region close to the black hole horizon. The black hole will capture all the trajectories with $x_{\rm min}<r_{+}$ -- this can be proved by considering the same process under time reverse, i.e., the particle moving from infinity towards the black hole with large $v_{\rm}$ and $x_{\rm min}<r_{+}$. Trajectories of such particles cannot 'jump over' the black hole, being captured by the black hole. 
For strongly accelerated charged particles the final $x$-coordinate after scattering in the vicinity of the black hole should be $x_{\rm min}\ge~r_{+}$. Because the outer Kerr black hole horizon $r_{+}\in\langle~1,2)$, we can expect strongly accelerated particles for parameters satisfying the relation $ 1<\cl/\cb<5 $.

The results of the scattering calculations are represented in Figure \ref{chaotic}. We clearly see how the regions of relatively regular character of the scattering process are mixed with quite chaotic regions of the scatter process. 
The charged particle acceleration by chaotic scattering (energy transmutation $\ce_{\rm z}\leftrightarrow\ce_{\rm 0}$) works well even in the \Schw{} spacetime combined with the uniform magnetic field. The black hole rotation is not required, although the energy transmutation can be more efficient in the rotating Kerr black hole spacetime, as demonstrated in Fig. \ref{chaotic}, since the gravitational potential of Kerr black holes is deeper than those of the \Schw{} black holes and the scatter could be more efficient. We can expect that the acceleration of particles by the transmutation effect can be even more efficient in the Kerr naked singularity spacetime \cite{Stu:1980:BAC:}, where the scattered charged particles cannot be captured by an event horizon and even particle collisions can be energetically more efficient in comparison to those occurring in the black hole spacetimes \cite{Stu-Sche:2013:CLAQG:}. 

There are continuous and discontinuous (chaotic) parts in the scattering function (escape velocity $\gamma_{\rm z}$) in dependence on the impact parameter (distance from equatorial plane $z_0$), see Fig. \ref{chaotic}. The existence of more then one unstable periodic orbit is responsible for occurrence of several regions of chaotic scattering, i.e., the discontinuous dependence of the resulting velocity on the initial conditions. The unstable periodic orbits are special orbits that are fixed in the spacetime, but they are extremely sensitive to changes of the initial conditions. 

For resulting ultra-relativistic acceleration, large initial charged particle energy $\ce$ and low orbital speed $u^\phi$, 
given by the condition $ 1<\cl/\cb<5 $,
are required. Such condition could be represented in the kinetic approach \cite{Cre-Stu-Tes:2013:PhysPlasm:}. Large velocities with $\gamma_{\rm z}\gg~1$ can be obtained for properly chosen initial conditions, but the question arise, if such conditions could be represented by realistic astrophysical conditions. 

The initial conditions enabling strong acceleration along the magnetic field lines and ultra-relativistic escape velocities of charged particles could be realized due to changing of ionization degree of ions following off-equatorial circular orbits that can exist in the field of magnetized black holes \cite{Kov-Stu-Kar:2008:CLAQG:,Kov-etal:2010:CLAQG:}; we shall study such a process in a future paper. Here we tackle the question of astrophysical relevance of the chaotic scattering acceleration model by considering very basic idea of ionization of neutral particles in Keplerian discs.

\begin{figure*}
\begin{center}\includegraphics[width=0.9\hsize]{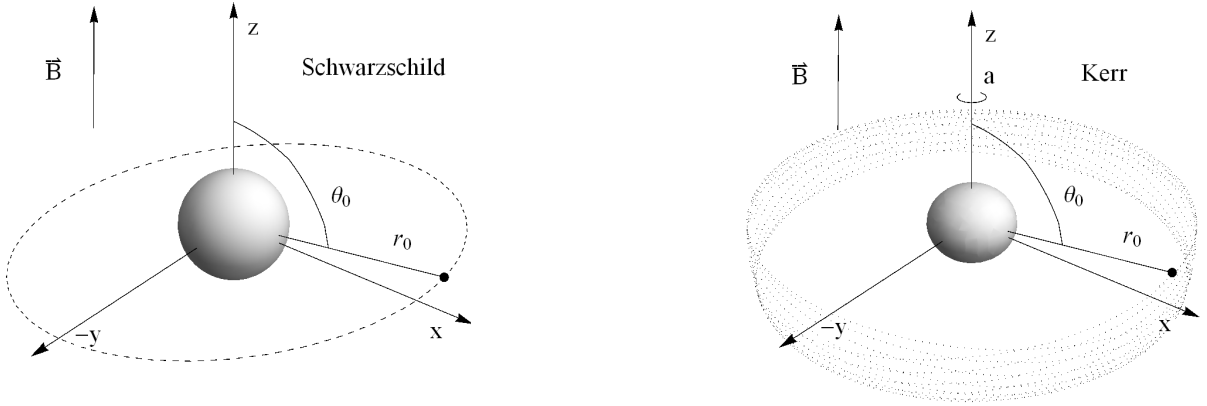}\end{center}
\caption{ \label{intro}
Initial orbits of test particles creating the neutral accretion disk. Due to spherical symmetry of the non-rotating \Schw{} black hole (left), the neutral particle trajectory lies always in central plane. The accretion disc inclination $\theta_0$ to the $z$ axis parallel with the uniform magnetic field vector, can be used to describe the magnetic field vector inclination to the accretion disc plane.
In the case of rotating Kerr black hole (right), we consider the neutral particle with initial inclination $\theta_0$ to be on the so called spherical orbit, given by (\ref{sphL}). Such scenario describe only situation when the magnetic field vector is perpendicular to the accretion discs plane - more general situation with some magnetic field vector inclination, will destroy the axial symmetry of charged particle dynamics around Kerr black hole immersed in the uniform magnetic field.
}
\end{figure*}


\section{Acceleration of ionized particles on circular geodesics}


We assume a thin accretion disk modelled by neutral test particles orbiting the central black hole along circular or quasi-circular geodesics. Ionization of a neutral particle, by a particle collision or irradiation, can change not only the test particle charge (from 0 to $q$),  but also its 4-momentum - the test particle can obtain some 'kick' during ionization. The ionization energy, i.e., the energy which is needed by an atom to loose one electron is around 10~eV, while the rest mass energy of the atom is around $10^9$~eV. If the ratio between the ionization and rest energy is very small, we can assume ionization with very small 'kick'. In fact, if we consider ionization of uncharged particles by an irradiation is assumed in the models of generation of profiled spectral lines \cite{Laor:1991:ApJ:,Fan-etal:1997:PASJ:,Sche-Stu:2009:GenRelGrav:}, then the energy of the kick of the ionizing photon is taken by the lost electron, while the ion (or proton) 4-momentum is not influenced by the ionization process, but it reflects the influence of the magnetic field. Therefore, the test particle mass and mechanical momenta before (\mJ) and after the ionization (\mD) are conserved 
\beq
m_{(\mJ)} = m_{(\mD)}, \quad p_{(\mJ)}^{\mu} = p_{(\mD)}^{\mu}. \label{ioniz}
\eeq
However, the motion constants start to be influenced by the electromagnetic field. 

Presented ionization scenario is obviously just some limit of more general (mechanical momenta and mass not conserving) ionization process, but it is quite realistic for the ionization by irradiation. The presented ionization model is identical to situation where the test particle is charged, but the external magnetic field is switched-off (non-existing), and suddenly the magnetic field is switched-on and the test particle starts to feel it.

The constants of the charged particle motion - energy $E$ and angular momentum $L$ are given by the generalized (canonical) momenta $\cp_{t}$ and $\cp_{\phi}$, see Eqs. (\ref{symE}-\ref{symL}). The conservation of mass and mechanical momenta during ionization, given by (\ref{ioniz}), requires the change of the particle specific energy $\ce$ and the specific angular momentum $\cl$ before and after ionization given by 
\beq
 \ce_{(\mD)} = \ce_{(\mJ)} - \tilde{q} A_{t}, \quad \cl_{(\mD)} = \cl_{(\mJ)} + \tilde{q} A_{\phi}. \label{ionCOND}
\eeq

The test particle mass and mechanical momenta remain the same during ionization (time $\tau_0$) and hence we can simply obtain the initial 4-velocity of the charged particle $u^{\mu}_{(\mD)}(\tau_0)$ from the neutral particle 4-velocity $u^{\mu}_{(\mJ)}(\tau_0)$ -- the particle speed is not changing during ionization, but after the ionization the charged particle  starts to feel the Lorentz force determined by the magnetic field. 

In general, the trajectory after ionization depends on the initial position 4-vector $t_0,r_0,\theta_0,\phi_0$ of the ionization event, the 4-velocity of the neutral particle at the ionization event, $u^t_0,u^r_0,u^\theta_0,u^\phi_0$, and also the parameters of the gravo-magnetic field - the black hole mass $M$ and spin $a$, and the magnetic field parameter $\cb$. The components of initial conditions $(t_0,r_0,\theta_0,\phi_0,u^t_0,u^r_0,u^\theta_0,u^\phi_0)$ are not independent, for example, the component $u^t_0$ can be calculated from the $H=0$ condition and we can choose the initial time $t_0$ arbitrary, fixing number of free initial components to six. 

We can reduce the set of the free initial components by assuming some astrophysically relevant initial conditions for the neutral test particles forming the Keplerian accretion disk. Assume a test particle on circular orbit with constant radius, $u^r_0=0$, close to the equatorial plane, with no vertical speed, $u^\theta_0=0$. Assuming a trajectory with constant radius, $r=r_0$, implies restrictions also on the particle angular momentum $\cl(r_0,\theta_0)$ and energy $\ce(r_0,\theta_0)$, i.e., on the 4-velocity components $u^{\phi}(r_0,\theta_0)$ and $u^{t}(r_0,\theta_0)$. 
There will be no dependence of the ionized trajectory on the initial coordinate $\phi_0$, if the ionization happens in equatorial plane $\theta_0=\pi/2$, and for $\theta_0\sim\pi/2$ the $\phi_0$-dependence will be insignificant. So we for simplicity select the ionization event to occur at $\phi_0=0$. The mass of the black hole can be simply put to be $M=1$, then the radius is expressed in units of black hole mass. 

For various gravo-magnetic field parameters $a,\cb$, we examine the ionization scenario only for the the two free parameters represented by the test particle initial coordinates $r_0, \theta_0$ of the ionization event.

\subsection{Ionization in the \Schw{} spacetime}

We consider a Keplerian accretion disk orbiting the \Schw{} black hole that consists from electrically neutral test particles following circular geodesics in the equatorial plane to which orthogonal are the magnetic field lines that are oriented parallel to the $z$-axis at infinity. The vector of the external uniform magnetic field can have in principle arbitrary inclination to the accretion disc plane, but here we assume for simplicity the orthogonal orientation that keeps the symmetry of the gravo-magnetic field. 
The spherical symmetry of the \Schw{} spacetime allows to rotate the whole neutral accretion disc, or some its part (ring), around the asymptotic $z$-axis, i.e., to introduce some small change of the inclination of the orbital plane of the disk (or its ring) relative to the magnetic field vector, see Fig. \ref{intro}. (left). 

The circular orbits ($r=r_0={\rm const.}$) of neutral test particles around non-rotating \Schw{} black holes are fixed to the central planes only, no matter what the initial inclination $\theta_0$ is. Angular momenta $\cl$ and energy $\ce$ for the inclined circular orbits are given by \cite{Wald:1984:book:}
\beq
 \cl = \frac{r \sin(\theta)}{\sqrt{r-3}}, \quad \ce = \frac{r-2}{\sqrt{r^2 - 3r}}. \label{CandLinSCHW}
\eeq

The $A_{t}$ component of electromagnetic four-vector potential is zero in the \Schw{} spacetimes, Eq. (\ref{AtApFlat}), hence we obtain for the ionization scenario by Eq. (\ref{ionCOND}) the following conditions
\beq
 \ce_{(\mD)} = \ce_{(\mJ)}, \quad \cl_{(\mD)} = \cl_{(\mJ)} + \tilde{q} A_{\phi}. \label{ionSCHW}
\eeq
Due to the vanishing $A_{t}$ component in the \Schw{} metric, only the angular momentum $\cl$ is changed during the ionization, while the particle energy $\ce$ remains constant. The energy of the neutral particles on the circular geodesic orbits, given by (\ref{CandLinSCHW}), is always $\ce_{(\mJ)}\leq~1$, but energy $\ce_{(\mD)}>1$ is needed for the ionized charged particle to escape to infinity along the $z$-axis (\ref{minE}). Therefore, no escape to infinity is possible after ionization (\ref{ionSCHW}) of neutral matter in Keplerian discs in the \Schw{} spacetime. However, the irradiational ionization can work for ions following the off-equatorial circular orbits around magnetized \Schw{} black hole \cite{Kov-etal:2010:CLAQG:}. 

When the magnetic field vector oriented asymptotically along the $z$-axis is orthogonal (or nealy orthogonal) to the plane of the orbiting ionized particle, the minima of the charged particle effective potential $V_{\rm eff}(r,\theta)$ are located in the equatorial plane $x$-$y$ \cite{Kol-Stu-Tur:2015:CLAQG:}. The angular momentum $\cl$ of the ionized particle has been changed by the ionization process (\ref{ionSCHW}) and after ionization the energy is not corresponding to a minimum of the effective potential, i.e., trajectory of the ionized particle will not be circular. Since the ionized particle cannot escape to infinity in the \Schw{} spacetime, $\ce_{(\mD)}\leq~1$, it can only be trapped in some bounded motion in vicinity of the black hole, or be captured by the black hole.

In the following calculations we assume the ionization event at the $\phi_0=0$ plane - such selection is taken just for simplicity. Due to the chaotic character of the equations of motions for the charged particles in the gravo-magnetic field, the particle trajectories with different initial conditions $\phi_0$ will be in principle different, but the $\phi_0=0$ selection reflects all the important properties of the ionization model. 
The particle angular momentum after ionization, $\cl_{(\mD)}$, will be given by
\beq
  \cl_{(\mD)} = r_0 \sin(\theta_0) \left( \frac{1}{\sqrt{r_0-3}} + \cb \right), \label{newLL}
\eeq
where we consider initial angular momentum $\cl_{(\mJ)}>0$. For positive magnetic field parameter, $\cb>0$, the term in parenthesis of (\ref{newLL}) has to be positive and hence we obtain new positive angular momentum $\cl_{(\mD)}>0$ giving thus a plus configuration according to our classification. For negative magnetic field parameter, $\cb<0$, the term in parenthesis of (\ref{newLL}) is positive only for small negative values of $\cb_{\rm b}<\cb<0$ giving a positive angular momentum $\cl_{(\mD)}>0$ after ionization, implying a minus configuration. The limiting value of the magnetic field parameter $\cb_{\rm b}$ reads
\beq
 \cb_{\rm b} = \frac{-1}{\sqrt{r_0-3}}.
\eeq
For sufficiently large negative values of the magnetic field parameter $\cb<\cb_{\rm b}$, the term in parenthesis of (\ref{newLL}) is negative, giving a negative new angular momenta $\cl_{(\mD)}<0$ - but because the magnetic field parameter is negative $\cb<0$, we obtain again a plus configuration. 
Clearly, the plus configurations, $\cl\cb>0$, are more likely created by the ionization process (\ref{ionSCHW}) for strong enough magnetic fields, $\cb<<0$, or large enough radius, $r_0>>r_{+}$. There is a preference to have a plus configuration after the ionization process. This fact implies large probability to obtain a large charged particle acceleration due to the ionization process, since for the plus configurations of our classification, the energy transmission process due to the chaotic scatter, $\ce_0\leftrightarrow\ce_{\rm z}$, works more efficiently in comparison to the minus configurations 
as demonstrated in the previous section. 

In the \Schw{} metric the ionized particle following originally a circular geodesic cannot escape to infinity, the capture by the black hole, or bound motion in its vicinity are the only options. The capturing of ionized particles by the black hole could lead to gradual disintegration of the Keplerian disc and fast accretion of its mass. This process depends on magnitude of the magnetic field and will be studied in detail in a future paper. 
If the charged particle is not captured, it will be bounded moving in some closed region around the black hole. The motion of such a bounded charged particle is in general chaotic. However, for small inclination of magnetic field vector to the axis of the accretion disc, $\theta_0\sim\pi/2$, the bounded motion will be regular, implying for the charged particles harmonic or quasi-harmonic oscillations with fundamental epicyclic frequencies that were calculated in \cite{Kol-Stu-Tur:2015:CLAQG:}. Therefore, the neutral particle ionization model (\ref{ionSCHW}) could also serve for explanation of quasi-periodic oscillations observed in the micro-quasars as shown in \cite{Kol-Stu-Tur:2015:CLAQG:}. 
By increasing the initial inclination angle $\theta_0$, the bounded oscillatory motion will be gradually changed from the regular harmonic motion to the quasi-harmonic motion where the epicyclic frequencies can still be relevant in the Fourier spectra, to the chaotic motion (with continuum spectrum) - for large initial inclination angles the motion will be fully chaotic. 

\subsection{Effect of the Kerr black hole rotation}

We keep the assumption of the magnetic field lines asymptotically parallel to the $z$-axis that is identical with the rotation axis of the Kerr black hole spacetime. The straightforward approach to the irradiation ionization scenario of a test particle from the electrically neutral Keplerian accretion disc in rotating Kerr spacetime means that we assume the neutral test particle to follow a circular Keplerian (geodesic) corotating orbit in the equatorial plane with the covariant specific energy $\ce$ and the specific axial angular momentum $\ce$ given by the standard relations \cite{Bar-Pre-Teu:1972:ApJ:}
\beq
 \ce=\frac{\frac{a}{r^{3/2}}-\frac{2}{r}+1}{\sqrt{\frac{2 a}{r^{3/2}}-\frac{3}{r}+1}}, \quad \cl = \frac{\frac{a^2}{r^{3/2}}-\frac{2 a}{r}+\sqrt{r}}{\sqrt{\frac{2 a}{r^{3/2}}-\frac{3}{r}+1}}. \label{ELeqplane}
\eeq

However, if we assume the irradiational ionization of a neutral particle following an equatorial circular geodesic with energy and angular momentum given by (\ref{ELeqplane}), with the magnetic field vector parallel with the symmetry $z$-axis, the trajectory of the ionized charged particle will stay in the equatorial plane. The coordinate $\theta=\pi/2$ remains constant and the motion is regular being effectively one dimensional. Such charged particle can only radially oscillate or be captured by the black hole. In order to have a possibility of off-equatorial motion that could be transmuted into the escaping motion of the charged particle, we have to assume a quasi-circular, off-equatorial epicyclic motion of the neutral particle and its ionization at a positions where their latitudinal coordinate $\theta_{0}\neq\pi/2$. 

\begin{figure*}
\includegraphics[width=\hsize]{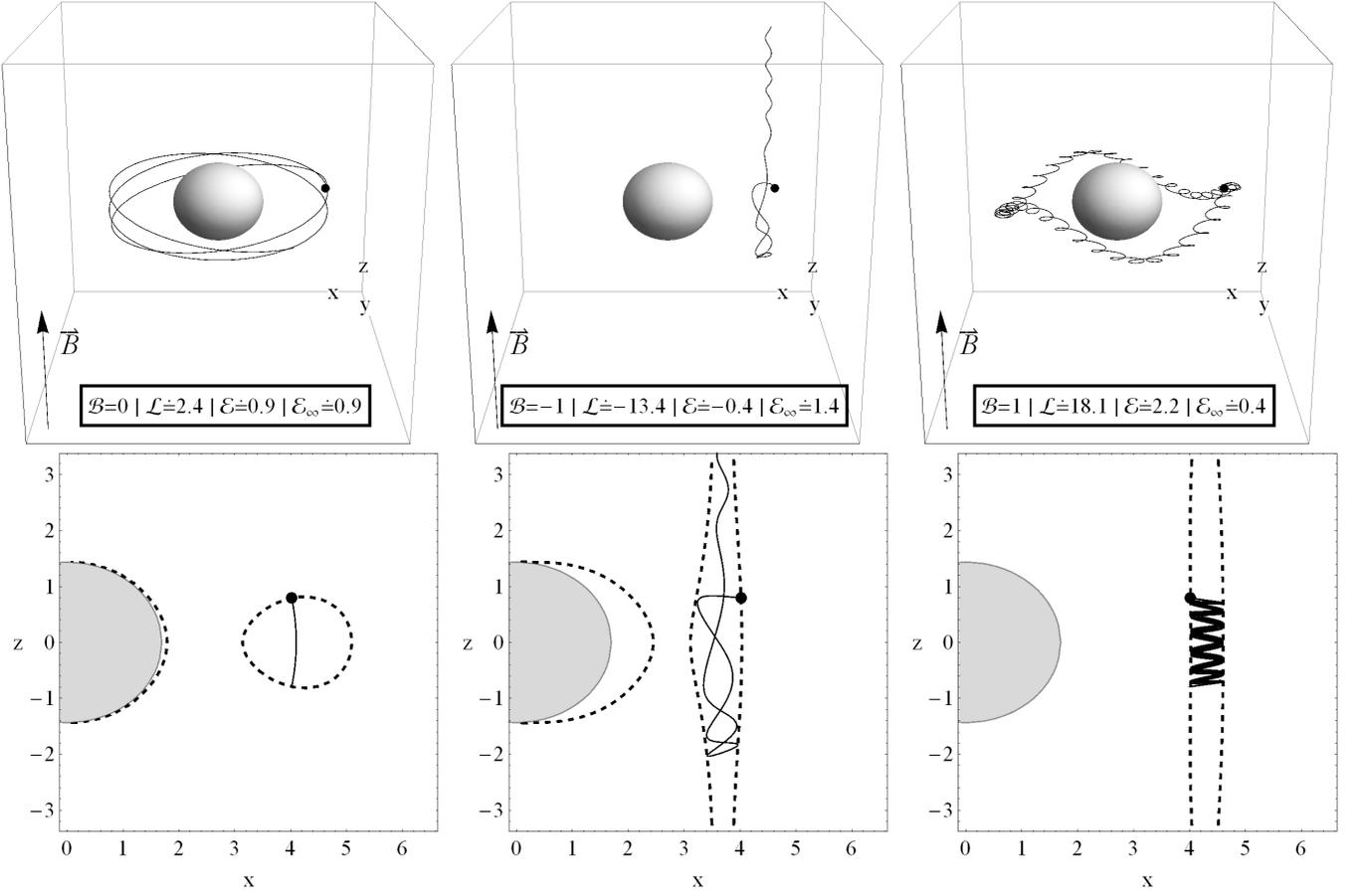}
\caption{
\label{ionKERR}
Ionization of Keplerian disc by irradiation. Neutral test particle from the accretion disc is orbiting the rotating Kerr black hole (left) and get ionized (\ref{ioniz}) in external  uniform magnetic field.
For negative value $\cb=-1$ of the magnetic field parameter, the ionized particle escapes along the magnetic filed lines (middle); for positive value $\cb=-1$ the ionized particle is periodically oscillating (right). Note that in the both cases, $\cb=\pm1$, the charged particle belongs to the plus $\cb\cl>0$ configuration, since such configuration is preferred by the ionization process (\ref{ioniz}) .
The neutral particle was initially on spherical orbit (\ref{sphL}) with constant radius $r_0=4$ and initial inclination $\theta_0=\pi/2-0.2$ around the Kerr black hole with spin $a=0.9$. The external uniform magnetic field vector is aligned with the axis of spacetime symmetry $z$ and the black hole electric charge $Q=0$.
}
\end{figure*}

We will consider here for simplicity a special character of the off-equatorial motion, namely the motion along spherical trajectories where the orbit radius $r_{0}$ remains constant, but the latitude $\theta$ of the motion changes, see Fig. \ref{intro}. The irradiational ionization event is then assumed at some $\theta_{0}\neq\pi/2$ enabling the escaping motion of the ionized particle. The energy $\ce$ and angular momentum $\cl$ of an electrically neutral particle following a spherical orbit are given by \cite{Sha:1987:SAL:}
\bea
\ce &=& \frac{1}{\sqrt{\SS}}\left( 1 - \frac{2 r}{\RS^2} + \frac{a \QS}{\RS^2 \sqrt{r}} \sin{\theta} \right) , \label{sphE}\\
\cl &=& \frac{1}{\sqrt{\SS}}\left( \frac{\QS (r^2+a^2)}{\sqrt{r}\RS^2} \sin(\theta) -\frac{2 a r}{\RS^2} \sin^2(\theta) \right) , \label{sphL}
\eea
where the $\RS$ function is defined by (\ref{RSaDD}) and the functions $\QS,\SS$ are given by the relations 
\bea
\QS &=& \sqrt{r^2 - a^2 \cos^2(\theta)}, \\
\SS &=& 1 - \frac{3r}{\RS^2} + \frac{2 a \QS}{\RS^2 \sqrt{r}} \sin(\theta) + \frac{a^2}{\RS^2 r} \cos^2(\theta) .
\eea
The equations (\ref{sphE}) and (\ref{sphL}) reduce to (\ref{CandLinSCHW}) for $a=0$; they reduce to the expressions for the equatorial Keplerian orbits (\ref{ELeqplane}) when $\theta=\pi/2$. 
The innermost stable spherical orbit (ISSO) of particles on the spherical orbits is implicitly given by the relation
\begin{widetext}
\bea
&& \left[ 2aQ \sqrt{r} \sin(\theta) -4r^2 +(r+1) R^2 \right]
\left[ Q^2 \left(R^2 \left(a^2-3 r^2\right)+4 r^2 \left(a^2+r^2\right)\right)-2 r^2 R^2 \left(a^2+r^2\right)-4 a Q^3 r^{3/2} \sin (\theta ) \right] \nn\\
&& + \left[4 Q^2 r^2-R^4\right] \left[Q-a \sqrt{r} \sin (\theta )\right] \left[Q \left(a^2+r^2\right)-2 a r^{3/2} \sin (\theta )\right] = 0. \label{sISCO}
\eea
\end{widetext}

\begin{figure*}
\includegraphics[width=\hsize]{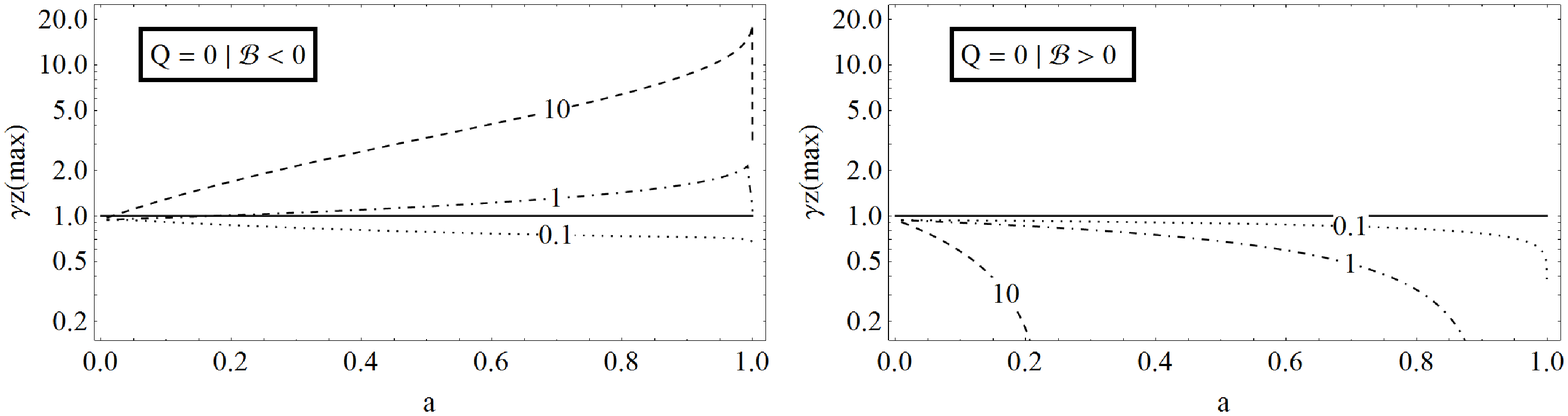}
\includegraphics[width=\hsize]{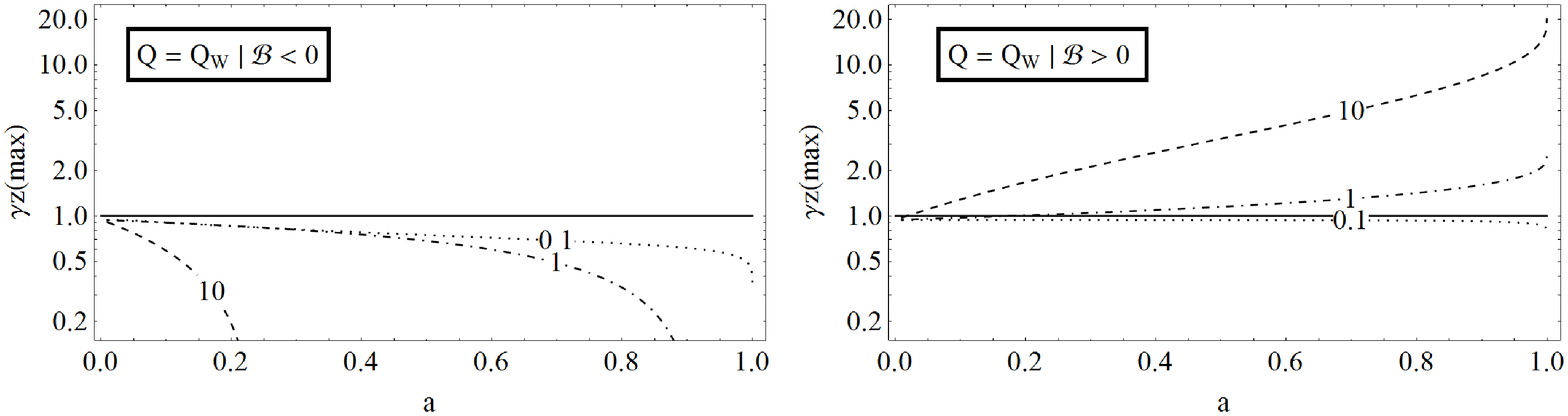}
\caption{
\label{maxGAM}
Maximal acceleration of ionized particles. We give limit on maximal possible charged particle gamma factor $\gamma_{\rm (z) max}$ given by the ratio: energy measured at infinity $\ce_{\infty}$ over energy needed to reach infinity $\ce_{0}$ (\ref{speedZmaxM}-\ref{speedZmaxP}) in dependence on the Kerr black hole spin $a$ for various values of magnetic field parameter $\cb$. 
Maximal possible gamma factor $\gamma_{\rm (z)max}$ must be $\gamma_{\rm (z)max}>1$ for the particle escape, values of $\gamma_{\rm (z)max}<1$ does not allow the motion of charged particle towards infinity along $z$ axis - the energetic boundary $\ce=V_{\rm eff}(x,y)$ is closed in the $z$ direction. The condition $\gamma_{\rm (z)max}=1$ is represent by the black line.
We consider the ionization (\ref{ionCOND}) to occur at the inner edge of accretion disc, located at $r_0=r_{\rm ISCO}$ (\ref{sISCO}) for initial inclination $\theta_0=\pi/2-0.1$. 
Both cases of neutral Kerr black hole with $Q=0$ and the black hole with the Wald charge $Q=Q_{\rm W}$ (\ref{WaldCharge}) are considered; numbers on the curves indicate the magnitude of the $\cb$ parameter.
}
\end{figure*}

\begin{figure*}
\includegraphics[width=\hsize]{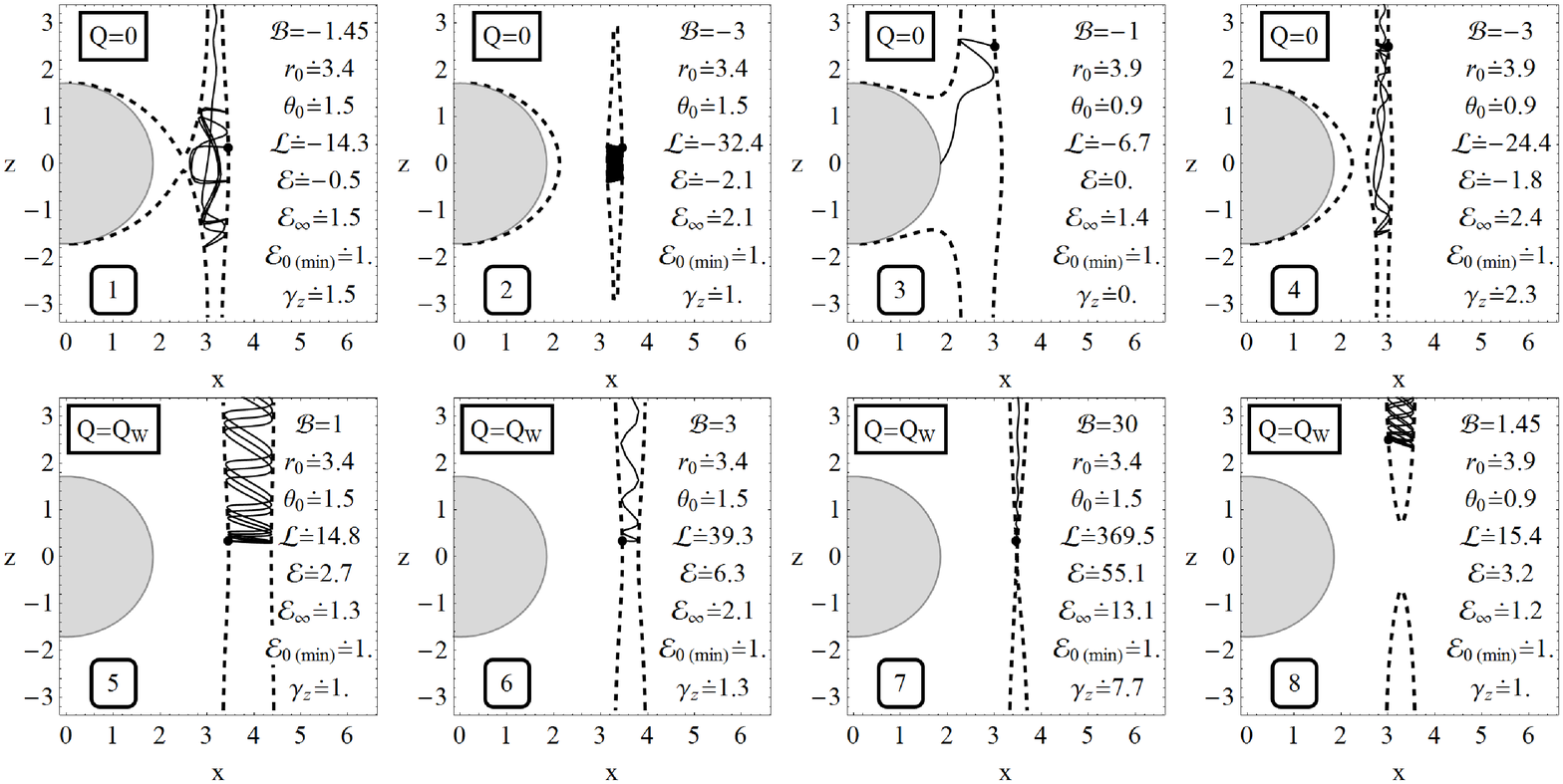}
\caption{
\label{ionISCO}
Irradiational ionization of Keplerian discs. Trajectories of charged particle ionized at the inner edge on neutral accretion disc around rotating Kerr black hole in external uniform magnetic field, given for characteristic values of magnetic field parameter $\cb$. As examples two representative initial positions of the ionization are used: close to the equatorial plane $\theta_0=\pi/2-0.1\doteq1.5, r_0\doteq~3.4, \cl_{(\mJ)}\doteq~2.6, \ce_{(\mJ)}\doteq~0.9$ and off the equatorial plane $\theta_0=\pi/2-0.7\doteq0.9, r_0\doteq~3.9, \cl_{(\mJ)}\doteq~2.1, \ce_{(\mJ)}\doteq~0.9$. The values of conserved energy $\ce_{(\mD)}$ and angular momenta $\cl_{(\mD)}$ after the ionization also with the $\gamma_{\rm z}$ calculated at $r=10^3$ are given in the figures.
For the neutral Kerr black hole ($Q=0$) we see the charged particle escaping to infinity with almost all possible gamma factor $\gamma_{\rm z}$ used (Fig. 1. and 4.); is oscillating close to equatorial plane (Fig. 2.) - initial angle $\theta_0$ is too small; is captured by black hole (Fig. 3.). 
For the Kerr black with Wald charge hole $Q=Q_{\rm W}$, we see the charged particle going repeatedly going away along $z$ axis and coming back close to black hole (Fig. 5. and 8.); is escaping with some large velocity along $z$ axis (Fig. 6. and 7.) - not all possible gamma factor $\gamma_{\rm z}$ is used. 
}
\end{figure*}

Typical examples of trajectories of ionized neutral test particles, initially orbiting on spherical orbits with radius $r_0\approx~r_{\rm ISSO}$ around a Kerr black hole immersed in the uniform magnetic field with field lines aligned with the spacetime axi-symmetry axis, can be found in Fig. \ref{ionKERR}. Contrary to the \Schw{} case with $a=0$, the energy shift $\ce_{(\mJ)}\rightarrow\ce_{(\mD)}$ given by Eq.(\ref{ionCOND}) and governed by the non-zero $A_{t}$ component of electromagnetic potential, allows the charged ionized particle to escape toward infinity along the $z$-axis. Of course, independence on the magnitude of the magnetic field $\cb$, rotation parameter $a$ of the Kerr spacetime, and the initial conditions, after irradiation, the ionized particle (\ref{ionCOND}) can be also captured by the black hole, or oscillate in some region in vicinity of the black hole horizon.

To obtain an astrophysically plausible situation, we have to consider the ionization event taking place near the inner edge of the neutral accretion disc. For simplicity we will assume the ionization event located at the innermost spherical orbit. This selection of radial coordinate, $r=r_{\rm ISSO}$, is completely arbitrary, and another value of radial coordinate at the ionization event $r>r_{\rm ISSO}$, giving similar results.

After ionization the charged particle can escape to infinity, if $\ce_{(\mD)}\geq\ce_{\rm min}$ -- see Eq.(\ref{minE}). The possibility of ionized charged particle to escape to infinity can be also expressed in terms of  the maximal possible Lorentz gamma factor along the $z$-axis, $\gamma_{\rm z(max)}=\ce_{\infty}/\ce_{\rm 0(min)}$. Due to Eqs. (\ref{speedZmaxP}-\ref{speedZmaxM}) the escaping particle must have $\gamma_{z}>1$. We have calculated the maximal possible gamma factor $\gamma_{\rm z(max)}$ after ionization of neutral particle located at the ionization event with $r_0=r_{\rm ISSO}$ and some latitudinal angle $\theta_0\sim\pi/2$ for various values of the magnetic field intensity parameter $\cb$, and in dependence on the black hole spin $a$;  the results are presented in Fig. \ref{maxGAM}.
We can see that a particle can escape to infinity after ionization governed by the conditions expressed by Eq.(\ref{ionCOND}) only for negative values of the magnetic field parameter $\cb<0$ in the case of uncharged Kerr black hole with $Q=0$, but it can escape for positive values parameter $\cb>0$, if the Kerr black hole carries the Wald charge $Q=Q_{\rm W}$. 

The condition $\gamma_{\rm z(max)}>1$ is just the necessary condition for escape -- the escape to infinity can occur with some lower speed, $1<\gamma_{\rm z}<\gamma_{\rm z(max)}$; $\gamma_{\rm z(max)}$ is just the limit on maximal possible velocity. The trajectories can be also captured by the black hole. As demonstrated in Fig. \ref{maxGAM}., quite large escape velocities along the $z$-axis can be obtained, if we assume the magnetic field parameter $\cb$ large enough. We can expect to observe the trajectories with the Lorentz gamma factors $\gamma_{\rm z}>5$ for the magnetic field parameter $\cb<10, Q=0$ and $\cb>10, Q=Q_{\rm W}$, if the black hole spin $a>0.7$. Some typical trajectories of the charged particles ionized by irradiation at the inner edge of the Keplerian accretion disc, $r_0=r_{\rm ISSO}$, are calculated and represented by the two-dimensional sections in Fig. \ref{ionISCO}. 

After the ionization the charged particle is not always escaping to infinity with the maximal velocity related to $\gamma_{\rm z(max)}$. It can escape to infinity with some smaller velocity, $\gamma_{\rm z} < \gamma_{\rm z(max)}$, it can stay near the black hole being oscillating near the equatorial plane, or it can be captured by the black hole, as demonstrated in Fig. \ref{maxGAM}. The main differences between the results of the ionization in the gravo-magnetic field with the uncharged black hole ($Q=0$) or with the black hole having the induced Wald charge ($Q=Q_{\rm W}$), is in the shape of the effective potential, which is wider in the equatorial plane for the $Q=0$ case as compared to the $Q=Q_{\rm W}$ case, see Fig. \ref{figVeff}. Although the energetic conditions on $\gamma_{\rm z(max)}$ are similar for both $Q=0$ and $Q=Q_{\rm W}$ cases, we can expect that in the $Q=Q_{\rm W}$ case the ionization process will be more efficient in expelling the charged particles to infinity due to the effective potential shape -- see Fig. \ref{ionISCO}. 

As already stated in the previous sub-section related to the case of the \Schw{} spacetime, the ionization mechanism governed by Eq. (\ref{ionCOND}) creates the charged particles almost exclusively in the plus configurations with $\cb\cl>0$, which is a positive effect as related to the acceleration of the charge particles to infinity. Clearly, the acceleration of ionized particles in the plus configurations with $\cb\cl>0$ is more effective because of the condition $\ce_{\rm 0(min)}=1$, as shown in Section \ref{velocitiesSEC}.

Detailed analysis of the trajectories of ionized particles and statistics of the ionization event leading to  capture/oscillations/escape in dependence on the gravo-magnetic configurations and the initial conditions in the Keplerian discs are left for a future work. In the present paper we restricted our attention to a clear demonstration of the possibility of charged particle escape with relativistic velocity due to the chaotic scattering enabling transmutation of energy modes near a Kerr black hole immersed in uniform magnetic field because of the ionization process by irradiation of innermost region of the Keplerian disc. The ionization process assumes simply the mechanical momentum conservation. 
 
We have shown that the presented mechanism of the particle acceleration could be relevant in real astrophysical process leading to creation of collimated relativistic jets. We have shown that the mechanism works well even for non-extreme black holes with spin $a\sim0.7$ and a non-extreme magnitude of the magnetic field $\cb\sim10$. Magnitude of the magnetic field parameter $\cb$ in physical units is given in Tab. \ref{tab1}. 

\begin{center}
\begin{table}[ht]
{\small
\hfill{}
\begin{tabular}{| @{\quad} c @{\quad} | c @{\quad} | c @{\quad} | c @{\quad} |} 
\hline
  $\cb$ & $B_{\rm e-}$ [mGs] & $B_{\rm p+}$ [Gs] & $B_{\rm Fe}$ [Gs]  \\
\hline \hline
 10 & 23 & 43 & 2373\\
\hline
\end{tabular}
}
\hfill{}
\caption{Magnitude of the magnetic field in physical units for parameter $\cb=10$ calculated for black hole with mass $M=10~M_{\odot}$. We present values for electron, $B_{\rm e-}$, proton, $B_{\rm p+}$, and partially ionized (one electron lost) iron atom, $B_{\rm Fe-}$, in Gauss units. 
For different value of magnetic field parameter $\tilde{\cb}$, just multiply the numbers in the table by $\tilde{\cb}/10$ factor; for more massive central object $\tilde{M}$, just multiply the numbers in the table by $10/\tilde{M}$ factor.
\label{tab1}
} 
\end{table}
\end{center}

\section{Conclusions}

Magnetized black holes, i.e., black holes immersed in an uniform magnetic field represent astrophysically relevant and interesting model for the charged test particle dynamic related to the accretion or excretion-jet phenomena. The charged test particle dynamics around magnetized black holes demonstrates interesting properties. 
\begin{itemize}
\item Non-linear equations of motion imply the charged particle chaotic dynamics.
\item In the case of magnetized black holes, off-equatorial circular orbits can exist.
\item Charged particles are allowed to escape along the magnetic field lines to infinity.
\end{itemize}

From the astrophysical point of view, high relevance is related to the energy transmutation effect in the combined gravitational and magnetic fields of magnetized black holes that enables transmission of energy between the linear translational mode $\ce_{\rm z}$ and the oscillatory mode $\ce_{0}$ of the motion of charged particles. The transmutation effect is reflected by the following points. 
\begin{itemize}
\item Chaotic dynamics close to the black hole horizon is responsible for the transmutational energy interchange, $\ce_{\rm z}\leftrightarrow\ce_0$; far away from the region of chaotic motion, in the asymptotically uniform magnetic field, both the energy modes, $\ce_{\rm z}$ and $\ce_0$, are independently conserved and no energy interchange is possible. 
\item This effect does not require black hole rotation, being purely 'mechanical' characteristics of the chaotic motion enabling interchange of the energy in the different modes. 
\item The energy transmutation effect enables ejection of charged particles from the region close to the equatorial plane of the magnetized black hole along the axis of symmetry with relativistic velocities, giving thus a possibility to create relativistic jets observed in active galactic nuclei and microquasars.
\item Existence of unstable periodic orbits in the region of chaotic motion implies the discontinuous dependence of the escaping velocities of charged particles on the initial conditions.
\end{itemize}

The acceleration process giving the relativistic escaping velocities at infinity and possibility of creation of relativistic jets can work well even in the process of irradiational ionization of originally electrically neutral particles following near-circular motion in Keplerian accretion discs. In the ionization process due to irradiation, the created protons or heavy ions feel no kick, but they could be accelerated by pure switch-on of the electromagnetic force due to the electric charge induced on the irradiated particle. The irradiational ionization implies the following results. 
\begin{itemize}
\item The charged particles occurring due to the irradiational ionization in the \Schw{} spacetime with the uniform magnetic field start to oscillate or are captured by the black hole -- they cannot escape to infinity along the $z$-axis, as in the spherically symmetric spacetimes energy of the ionized matter remains constant and it corresponds to the bounded motion for the originally electrically neutral particle.
\item Escape to infinity along the magnetic field lines is possible for ionized particles in the field of magnetized rotating Kerr black holes with magnetic field lines oriented parallely to the symmetry axis of the spacetime, since the energy of the charged particle can be increased after the magnetic field switch-on. This effect can be relevant also for mediate and small values of the black hole spin. 
\item The irradiational ionization process prefers the states with coincidence of orientation of the particle angular momentum and the magnetic field vector, allowing for efficiently accelerated motion of the ionized particles along the magnetic field lines, giving an interesting new mechanism for creation of collimated relativistic jets around  rotating Kerr black holes with arbitrary value of the dimensionless spin $a$. 
\end{itemize}

Details of the transmutation effect causing acceleration of the irradiational ionized matter, following originally near-equatorial motion around a magnetized Kerr black hole, in the direction of the magnetic field lines will be studied in a future paper. However, attention has to be focused also for application of this model of ionization to describe the accretion disc disintegration and absorption (trajectories captured by the black hole), or generation of the charged particle quasi-harmonic oscillations used to explain the quasi-periodic oscillations observed in microquasars (bounded trajectories).

\section*{Acknowledgments}

The authors would like to express their acknowledgments for the Institutional support of the Faculty of Philosophy and Science of the Silesian University in Opava and the Albert Einstein Centre for Gravitation and Astrophysics supported by the Czech Science Foundation grant No.~14-37086G.



\def\prc{Phys. Rev. C}
\def\pre{Phys. Rev. E}
\def\prd{Phys. Rev. D}
\def\jcap{Journal of Cosmology and Astroparticle Physics}
\def\apss{Astrophysics and Space Science}
\def\mnras{Mon. Not. R. Astron Soc.}
\def\apj{The Astrophysical Journal}
\def\aap{Astron. Astrophys.}
\def\actaa{Acta Astronomica}
\def\pasj{Publications of the Astronomical Society of Japan}
\def\apjl{Astrophysical Journal Letters}
\def\pasa{Publications Astronomical Society of Australia}
\def\nat{Nature}
\def\physrep{Phys. Rep.}
\def\araa{Annu. Rev. Astron. Astrophys.}
\def\apjs{The Astrophysical Journal Supplement}
\def\aapr{The Astronomy and Astrophysics Review}

\end{document}